\documentclass[twocolumn,epjc3]{svjour3}
\usepackage{graphics}
\usepackage{microtype}
\usepackage{bbm}
\usepackage{amsmath}
\usepackage{mathtools}
\usepackage{caption}
\usepackage{subcaption}
\usepackage{amssymb}
\usepackage{mathrsfs}       
\usepackage[pdftex]{graphicx}
\usepackage{pgfplots}
\pgfplotsset{width=10cm,compat=1.9}
\usepackage{xargs}       
\usepackage{wasysym}
\usepackage{engrec}
\usepackage{enumerate}
\usepackage{appendix}
\usepackage{empheq}
\usepackage{float}
\usepackage{stmaryrd}
\usepackage[parfill]{parskip}
\usepackage[pdftex,breaklinks]{hyperref}
\usepackage{titletoc}
\usepackage{makecell}
\usepackage{lscape}
\usepackage{algorithm}
\usepackage{algorithmicx}
\usepackage{tensor}
\usepackage{algpseudocode}
\usepackage{blkarray}
\usepackage{multirow}
\usepackage{bigstrut}
\usepackage{booktabs}
\usetikzlibrary{plotmarks}

\usepackage[numbers,sort&compress]{natbib}

\newcommand{\jhat}[1]{\hat{#1}}

\newcommand{\elem}[2]{$^{#1}${#2}}

\newenvironment{customlegend}[1][]{%
    \begingroup
    \csname pgfplots@init@cleared@structures\endcsname
    \pgfplotsset{#1}%
}{%
    \csname pgfplots@createlegend\endcsname
    \endgroup
}%
\def\addlegendimage{\csname pgfplots@addlegendimage\endcsname}

\newcommand{\ai}{\textit{ab initio}}

\newcommand{\comment}[2]{\textcolor{red}{\textbf{#1:} #2}}

\definecolor{mygreen}{RGB}{21, 158, 10}

\begin{document}
\allowdisplaybreaks
\title{Impact of correlations on nuclear binding energies}
    \subtitle{\textit{Ab initio} calculations of singly and doubly open-shell nuclei}
\author{A. Scalesi\thanksref{ad:saclay} 
\and T. Duguet\thanksref{ad:saclay,ad:kul} 
\and P. Demol\thanksref{ad:kul}
\and M. Frosini\thanksref{ad:des}
\and V. Som\`a\thanksref{ad:saclay} 
\and A. Tichai\thanksref{ad:ger1,ad:ger2,ad:ger3} 
}

\institute{
\label{ad:saclay}
IRFU, CEA, Universit\'e Paris-Saclay, 91191 Gif-sur-Yvette, France 
\and
\label{ad:kul}
KU Leuven, Department of Physics and Astronomy, Instituut voor Kern- en Stralingsfysica, 3001 Leuven, Belgium 
\and
\label{ad:des}
CEA, DES, IRESNE, DER, SPRC, LEPh,
13115 Saint-Paul-lez-Durance, France
\and
\label{ad:ger1}
Technische Universit\"{a}t Darmstadt, Department of Physics, 64289 Darmstadt, Germany
\and
\label{ad:ger2}
ExtreMe Matter Institute EMMI, GSI Helmholtzzentrum f\"{u}r Schwerionenforschung GmbH, 64291 Darmstadt, Germany
\and
\label{ad:ger3}
Max-Planck-Institut f\"{u}r Kernphysik, 69117 Heidelberg, Germany
}


\maketitle
%
%
\begin{abstract}
A strong effort will be dedicated in the coming years to extend the reach of \ai{} nuclear-structure calculations to heavy doubly open-shell nuclei. In order to do so, the most efficient strategies to incorporate dominant many-body correlations at play in such nuclei must be identified. With this motivation in mind, the present work analyses the step-by-step inclusion of many-body correlations and their impact on binding energies of Calcium and Chromium isotopes.  

Employing an empirically-optimal Hamiltonian built from chiral effective field theory, binding energies along both isotopic chains are studied via a hierarchy of approximations based on polynomially-scaling expansion many-body methods. More specifically, calculations are performed based on (i) the spherical Hartree-Fock-Bogoliubov mean-field approximation plus correlations from second-order Bogoliubov many-body perturbation theory or Bogoliubov coupled cluster with singles and doubles on top of it, along with (ii) the axially-deformed Hartree-Fock-Bogoliubov mean-field approximation plus correlations from second-order Bogoliubov many-body perturbation theory built on it. The corresponding results are compared to experimental data and to those obtained via valence-space in-medium similarity renormalization group calculations at the normal-ordered two-body level that act as a reference in the present study.

The spherical mean-field approximation is shown to display specific shortcomings in Ca isotopes that can be understood analytically and that are efficiently corrected via the consistent addition of low-order dynamical correlations on top of it. While the same setting cannot appropriately reproduce binding energies in doubly open-shell Cr isotopes, allowing the unperturbed mean-field state to break rotational symmetry permits to efficiently capture the static correlations responsible for the phenomenological differences observed between the two isotopic chains.

Eventually, the present work demonstrates that polynomially-scaling expansion methods based on unperturbed states that possibly break (and restore) symmetries constitute an optimal route to extend \ai{} calculations to heavy closed- and open-shell nuclei.
\end{abstract}

\section{Introduction}
\label{intro}

Predictions based on \ai{} nuclear structure calculations are currently moving to heavier systems~\cite{Hu2021lead,Miyagi2021,Hebeler2023,Tichai:2023epe,Arthuis2024} and/or doubly open-shell nuclei~\cite{Novario:2020kuf,Hagen:2022tqp,Frosini:2021ddm,Frosini:2024ajq}. One ambition of such developments is to efficiently capture the dominant many-body correlations at play.  Qualitatively speaking, many-body correlations separate into two different categories. The first category concerns so-called {\it dynamical} correlations carried by all nucleons and delivering the bulk of the correlation energy. Dynamical correlations are well captured by a sum of many low-rank elementary, e.g. particle-hole, excitations out of a well-chosen unperturbed state. The second category concerns so-called {\it static} correlations that strongly impact the ground-state of open-shell nuclei and are driven by the valence nucleons. While being largely subleading, static correlations vary quickly with the number of valence nucleons and, as such, strongly impact differential quantities as well as spectroscopic observables. Such correlations can be efficiently captured via an optimal choice of the unperturbed state~\cite{Tichai:2020dna,Frosini:2021fjf}.

In this context, the present work wishes to analyse  the impact of the step-by-step inclusion of many-body correlations on binding energies and associated differential quantities, i.e. first- and second-order derivatives with respect to the (even) neutron number, while following different possible strategies to do so. The study is conducted along neighboring Calcium ($Z=20$) and Chromium ($Z=24$) isotopic chains spanning a large range of (even) neutron numbers from $N=12$ till $N=50$. Most of Ca isotopes are of singly open-shell character whereas most of Cr isotopes are of doubly open-shell character. Comparing the behavior of binding energies along these isotopic chains allows one to illustrate the roles played by static and dynamical correlations in the two types of nuclei and the capacity of \ai{} many-body methods to efficiently capture them by employing an optimal formulation. In order to control how some of the identified features depend on the nuclear mass, additional calculations are performed along the Tin ($Z=50$) isotopic chain from $N=50$ till $N=82$.

The paper is organized as follows. Section~\ref{setup} briefly characterises the numerical calculations performed in the present study. In Sec.~\ref{sMeanfield}, the results obtained at the spherical mean-field level are analysed, pointing to specific deficiencies that need to be remedied by the addition of correlations. In Sec.~\ref{sBeyonMeanfield}, low-order dynamical correlations on top of the spherical mean-field are proven to correct all such shortcomings to a high degree in Ca isotopes. In Sec.~\ref{dMeanfieldsyst}, the inclusion of static correlations either via a complete diagonalization in the valence space or via the explicit breaking of rotational symmetry is shown to be critical to obtain an equally good description of Cr isotopes. The paper is complemented by an appendix in which semi-analytical formulae are derived to provide a more intuitive understanding of the numerical results.

\begin{figure*}
    \centering
    \includegraphics[width=1.0\textwidth]{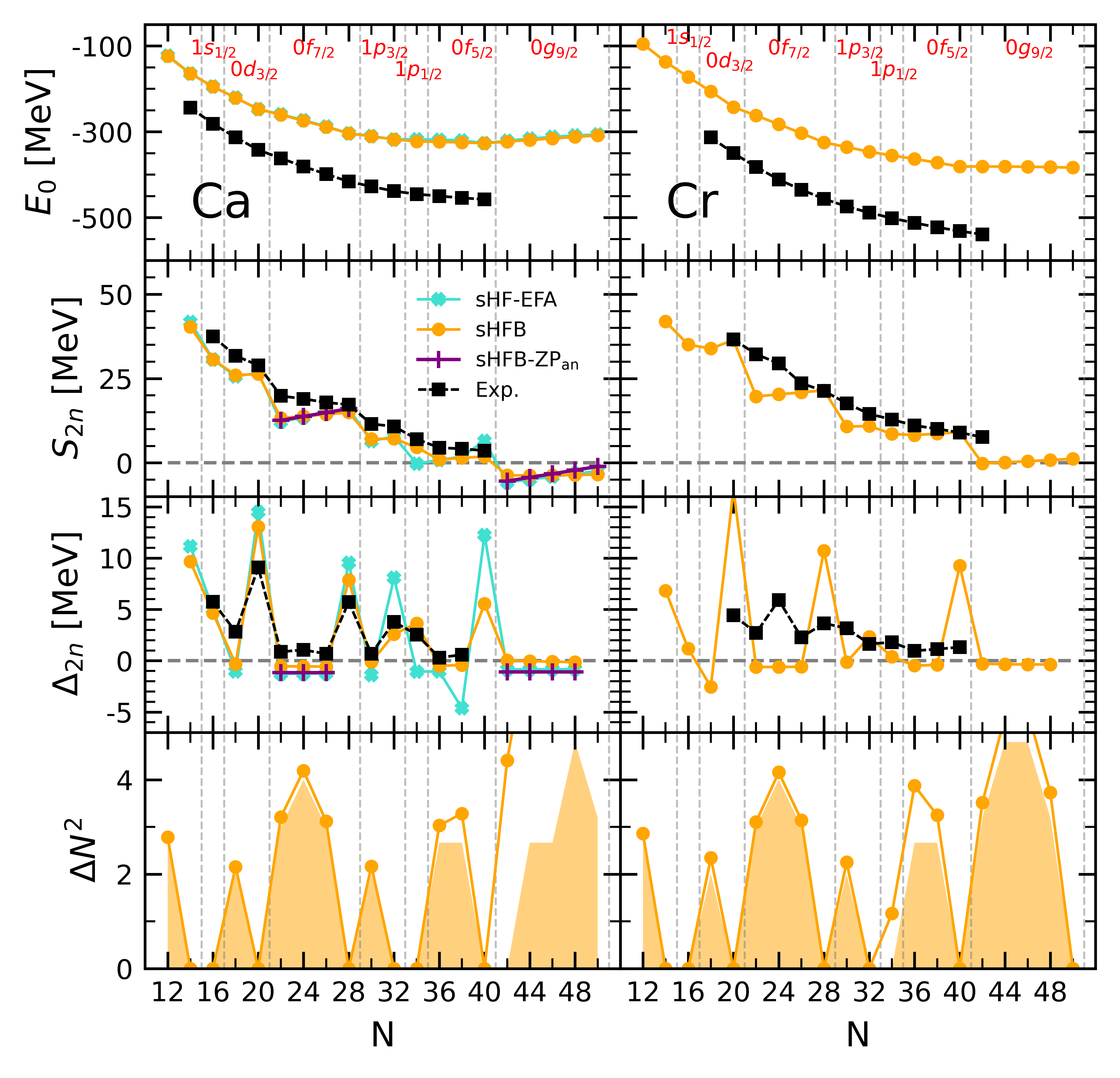}
    \caption{Systematics along Ca (left) and Cr (right) isotopic chains. First line: sHFB absolute binding energies against experimental data. Second (third) line: sHFB and sHF-EFA two-neutron separation energy (two-neutron shell gap) against experimental data. Between $^{42}$Ca and $^{48}$Ca (0f$_{7/2}$ shell) as well as between $^{62}$Ca and $^{70}$Ca (0g$_{9/2}$ shell), sHFB-ZP semi-analytical results are also shown. Fourth line: sHFB neutron-number variance against the minimal possible variance in sHFB calculations~\cite{Duguet:2020hdm}. The sequence of the underlying neutron canonical shells are also displayed.}
    \label{sHFB}
\end{figure*}

\section{Numerical calculations}
\label{setup}

\textit{Ab initio} many-body calculations are carried out employing a one-body spherical harmonic oscillator basis characterized by the frequency $\hbar\omega=12$~MeV. All states up to $e_{\!_{\;\text{max}}}\equiv\text{max}(2n+l)=12$ are included, with $n$ the principal quantum number and $l$ the orbital angular momentum. The representation of three-body operators is further restricted by only employing three-body states  up to $e_{\!_{\;\text{3max}}}=18 \, (24)$ in Ca and Cr (Sn) isotopes.

Calculations are performed using the EM~1.8/2.0 Hamiltonian from Ref.~\cite{Hebe11fits} containing  two-nucleon (2N) and three-nucleon (3N) interactions originating from chiral effective field theory ($\chi$EFT).
The 3N interaction is approximated via the rank-reduction method developed in Ref.~\cite{Frosini:2021tuj}. This particular Hamiltonian is employed because it is empirically known to give an excellent reproduction of binding energies in the mid-mass region~\cite{Stroberg:2021qu}.

The present study is based on three complementary expansion many-body methods. First, the spherical Hartree-Fock Bogoliubov (sHFB) mean-field approximation plus second-order Bogoliubov many-body perturbation theory (sBMBPT(2)) correction~\cite{Tichai18BMBPT,Arthuis:2018yoo} is employed. As a non-perturbative complement to sBMBPT(2), spherical Bogoliubov coupled cluster with singles and doubles (sBCCSD)~\cite{Sign14BogCC,Tichai:2023epe} calculations are also carried out. Third comes the axially-deformed Hartree-Fock Bogoliubov (dHFB) mean-field approximation plus second-order Bogoliubov many-body perturbation theory (dBMBPT(2)) correction~\cite{Frosini:2021tuj,Frosini:2024ajq} .

Available valence-space in-medium similarity renormalization group (VS-IMSRG(2)) results in Ca and Cr isotopes~\cite{Stroberg:2021qu} based on the same Hamiltonian\footnote{The numerical setting is slightly different given that these calculations employ $e_{\!_{\;\text{max}}}=14$ and $e_{\!_{\;\text{3max}}}=16$ while extrapolations in $e_{\!_{\;\text{max}}}$ are performed to obtain infrared convergence~\cite{Furnstahl:2014hca}. The effects of 3N interactions between valence nucleons is captured via the ensemble normal-ordering of Ref.~\cite{Stroberg:2016ung}.} are presently employed as a reference given that static and dynamical correlations generated within the valence space are accounted for to all orders via the diagonalization of the associated effective Hamiltonian. Notice that calculations along complete Ca and Cr isotopic chains require a reset of the valence space below $N=20$ and above $N=40$. The data presently employed correspond to the choice of valence spaces delivering the most optimal results~\cite{Stroberg:2021qu}.

\section{Spherical mean-field approximation}
\label{sMeanfield}

The baseline of more advanced treatment based on a many-body expansion is given by the mean-field approximation restricted to spherical symmetry~\cite{Hergert20}. Because the present study targets open-shell systems, the minimal version presently considered is given by sHFB that can naturally capture pairing correlations via the breaking of U(1) symmetry associated with particle-number conservation~\cite{RiSc80}.

\subsection{Ca chain}
\label{sMeanfieldsyst}

Systematic sHFB results along the Ca isotopic chain are displayed in the left panels of Fig.~\ref{sHFB}. In the first line, one observes that sHFB calculations significantly underbind experimental data, e.g. by more than $100$\,MeV in $^{48}$Ca, in a way that increases with neutron excess. Such a quantitative defect is expected from a mean-field approximation in the context of \ai{} calculations. Indeed, while static neutron-neutron pairing correlations are incorporated within sHFB, one is missing dynamical correlations whose inclusion account for a significant fraction of the binding energy~\cite{Hergert20,Tichai:2020dna,Frosini:2021ddm,Duguet:2022zup}. 

The evolution of binding energies can be scrutinized via the two-neutron separation energy
\begin{align}
    S_{2n}(N,Z) \equiv E(N-2,Z) - E(N,Z)
\end{align}
displayed in the second line of Fig.~\ref{sHFB}. Because $S_{2n}(N,Z)$ is a first derivative of the binding energy $E(N,Z)$ with respect to (even) neutron number, the large offset seen in the first line has disappeared. Eventually, the $S_{2n}$ from sHFB slightly underestimate experimental data overall such that adding dynamical correlations is expected to correct for this quantitative discrepancy. 

The main characteristics of the experimental $S_{2n}$, i.e. the sudden drops at $N=20$ and $28$, and to a lesser extent at $N=32$ and $34$, as well as the smooth evolution in between, are well accounted for by sHFB results. However, crucial differences are revealed upon closer inspection. First, the amplitude of the drops at $N=20$ and $28$ is too large and the trend in between, i.e. while filling the 0f$_{7/2}$ shell, is qualitatively wrong. Correlated with the too large drop at $N=20$, the $S_{2n}$ value in $^{42}$Ca is significantly too low. Further adding neutrons, $S_{2n}$ increases linearly throughout the 0f$_{7/2}$ shell instead of decreasing linearly as for experimental data\footnote{While the same patterns are at play when going through $^{48}$Ca, the size of the 1p$_{3/2}$ shell is too small to make the rising slope of sHFB results really visible. The highly degenerate 0g$_{9/2}$ shell between $^{62}$Ca and $^{70}$Ca is more favorable in this respect even though the slope of the sHFB results is actually zero in this case. These nuclei are anyway predicted to be unbound and there is no experimental data yet to be confronted with.}. 

Given that $S_{2n}(N,Z)$ is the first derivative of the binding energy, the patterns identified above relate to specific features of the binding energies that could not be fully appreciated in the first line of Fig.~\ref{sHFB} due to the large scale employed. The fact that $S_{2n}$ evolves linearly with the number $a_v$ of nucleons in the valence shell for both sHFB results and experiment data implies that $E(N,Z)$ is essentially quadratic in between two closed-shell isotopes. The fact that $S_{2n}$ starts from a too low value in sHFB calculations in the open-shell relates to the fact that the linear decrease of $E(N,Z)$ is not pronounced enough such that the difference to the data increases  throughout the shell. Finally, the fact that $S_{2n}$ is rising linearly instead of decreasing linearly indicates that the sHFB energy is concave instead of being convex. 

\begin{figure}
    \centering
    \includegraphics[width=0.5\textwidth]{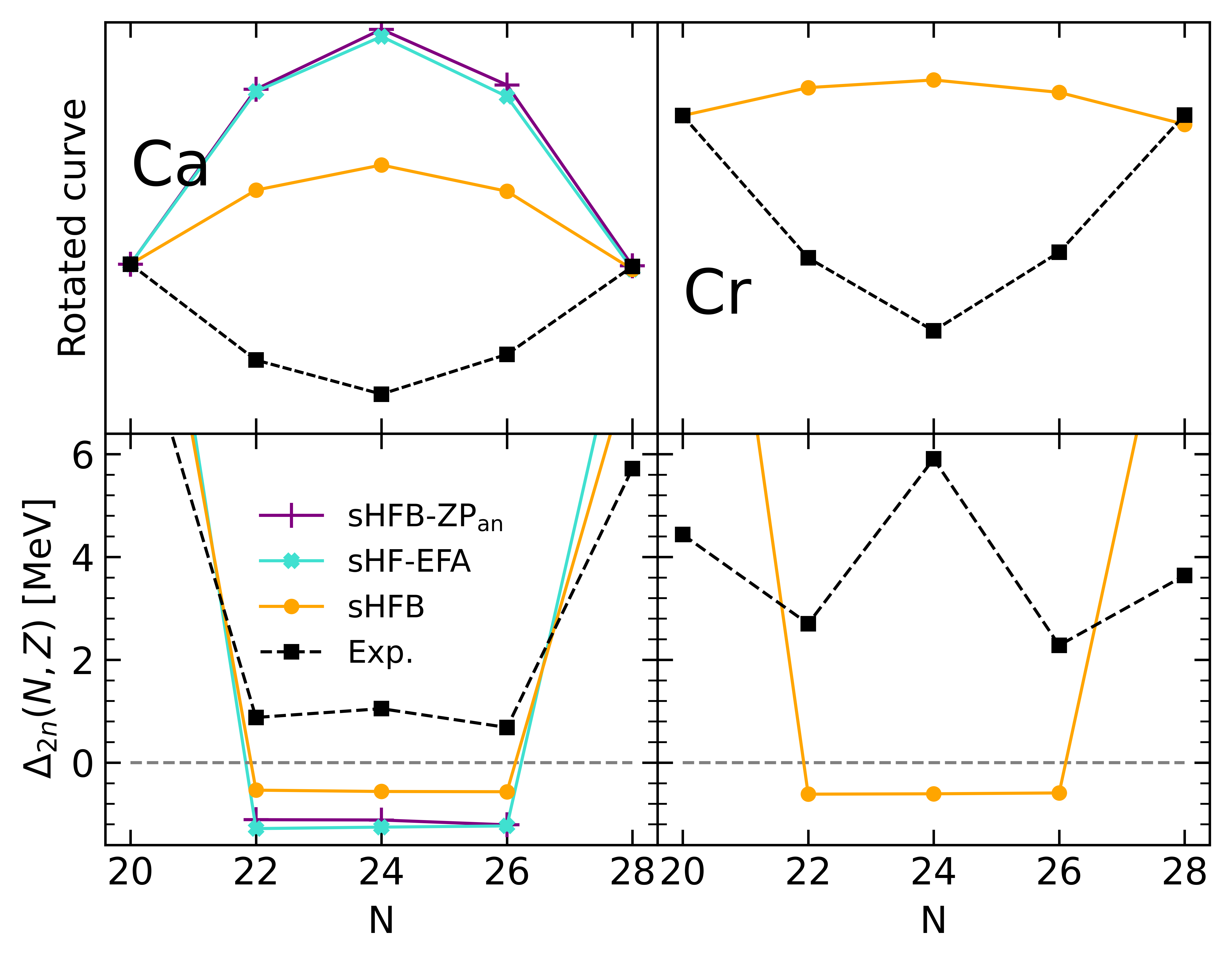}
    \caption{Energy curvature of Ca (left) and Cr (right) isotopes between $N=20$ and $N=28$ (0f$_{7/2}$ shell). Upper panels: energies rescaled to $N=20$ and $N=28$ (see text for details). Bottom panels: two-neutron shell gap $\Delta_{2n}$. Experimental data are compared to sHFB results. For Ca isotopes, results from sHF-EFA and sHFB-ZP semi-analytical results are also shown.}
    \label{sHFBcurvature}
\end{figure}

These characteristics can be pinned down quantitatively by looking at the third line of  Fig.~\ref{sHFB} displaying the so-called two-neutron shell gap
\begin{align}
    \Delta_{2n}(N,Z) \equiv  S_{2n}(N,Z) - S_{2n}(N+2,Z) \, .
    \label{eq:Delta2n}
\end{align}
Whenever $\Delta_{2n}$ displays a sudden increase, the amplitude of the spike provides an empirical measure of the extra stability associated with a mean-field picture of a closed-shell nucleus displaying a large Fermi gap. Otherwise, $\Delta_{2n}$ is linked to the second derivative, i.e. the curvature, of the smoothly evolving binding energy (see~\ref{app2ndderivative} for details). 

The left panel displaying $\Delta_{2n}$ in Fig.~\ref{sHFB} confirms the two patterns identified above. First, the amplitude of the spikes at $N=20$ and $N=28$ are too large by $4.0$ and $2.1$\,MeV, respectively\footnote{Contrarily, the sudden increase is correctly reproduced for $N=32$ and $N=34$. It seems that the larger the over-stability in the data, i.e. the more pronounced the magic character of the isotope, the larger the sHFB exaggeration.}. Second, the essentially constant character of the experimental $\Delta_{2n}$ between $^{42}$Ca and $^{48}$Ca is well captured by sHFB calculations but the associated value is negative instead of positive, i.e. to a very good approximation the sHFB energy is indeed quadratic with the number of valence nucleons $a_v$ but it is concave instead of being convex\footnote{Between $^{62}$Ca and $^{70}$Ca, where there is no experimental data, $\Delta_{2n}$  is constant but actually null such that the sHFB energy is rather linear with $a_v$.}.

Eventually, the issue associated with the curvature of the energy can be even better appreciated from the left panels of Fig.~\ref{sHFBcurvature} focusing on the isotopes between $N=20$ and $N=28$. While the bottom panel shows $\Delta_{2n}$, the upper panel displays the total energy rescaled to $N=20$ and rotated around that point such that the value at $N=28$ is aligned with it. This effectively removes the overall shift between the different curves along with the linear trend between the two closed-shell isotopes. Both panels make clear that, while experimental energies of Ca isotopes are essentially quadratic and convex between two closed-shell isotopes, sHFB calculations generate a quadratic dependence of energies whose curvature carries the wrong sign.

\subsection{Analytical investigation}
\label{sMeanfieldanalytic}

The wrong qualitative behavior of the sHFB energy along semi-magic chains was already visible in past calculations~\cite{Tichai18BMBPT,Soma:2020dyc,Tichai:2023epe} based on different chiral Hamiltonians. It seems to indicate that this behavior is deeply rooted into the spherical mean-field approximation based on realistic nuclear Hamiltonians. This expectation can in fact be confirmed analytically as demonstrated below.

In order to proceed, one must first make a crucial observation thanks to the results shown on the last line of Fig.~\ref{sHFB} comparing the neutron-number variance in the sHFB calculation to the {\it minimal} variance obtained in the zero-pairing limit of sHFB theory (sHFB-ZP)~\cite{Duguet:2020hdm}. As already noticed~\cite{Soma:2020dyc}, chiral Hamiltonians typically generate only little static pairing at the mean-field level\footnote{In an \textit{ab initio} setting, pairing properties such as the odd-even mass staggering are expected to largely originate from (i.e. required to account for) higher-order processes  associated with the exchange of {\it collective} medium fluctuations between paired particles~\cite{Barranco04,Gori05,Pastore08,Idini11}. Achieving a quantitative description of pairing properties from first principles constitutes a major challenge for \ai{} nuclear structure theory~\cite{Soma:2020dyc}.}, i.e. the computed neutron-number variance is indeed very close to the minimal variance in most open-shell isotopes, except in $^{56,58}$Ca and for nuclei in the continuum. As visible from the left panels of Fig.~\ref{sHFB}, this is confirmed by the proximity of sHFB results to those obtained from spherical Hartree-Fock calculations performed within the equal-filling approximation~\cite{Perez-Martin:2008dlm} (sHF-EFA) that do not include pairing correlations by construction. Results are indeed very close overall, except in $^{56,58}$Ca (0f$_{5/2}$ shell) where sHFB better reproduces experimental values for $S_{2n}$ and $\Delta_{2n}$. As for the curvature within open-shells, the left panels of Fig.~\ref{sHFBcurvature} reveals that the curvature of sHF-EFA results also carries the wrong sign but is such that the concavity is even more pronounced than for sHFB, i.e. the weak pairing correlations present within the 0f$_{7/2}$ shell in sHFB do improve the situation compared to the case where pairing would indeed be strictly zero.

Based on this observation, the sHFB energy of an open-shell nucleus relative to the closed-shell (CS) core\footnote{The following considerations can be meaningfully applied only to \textit{singly} open-shell nuclei, since they rely on the existence of a spherical core and highly-degenerate shells on top of it.} can be, to a good approximation, expressed analytically as a function of $a_v$ and of specific 2N and 3N interaction matrix elements within sHFB-ZP and sHF-EFA. Both cases are worked out in details in~\ref{analytical}. Since both variants provide almost identical numerical results, only the simpler sHF-EFA expressions are reported here whereas the complete set of formulae valid in sHFB-ZP can be found in~\ref{analytical}. 

Canonical single-particle states $k\equiv (n_k,l_k,j_k,m_k,\tau_k)$ diagonalizing the one-body density matrix $\rho^\text{sHF-EFA}$ gather in shells carrying degeneracy $d_k \equiv 2j_k+1$ characterized by the single-particle energies $\epsilon_{k} = \epsilon_{\breve{k}}$ (see Eq.~\ref{speEFAbulk} below) where $\breve{k}\equiv (n_k,l_k,j_k,\tau_k)$. For a system with A (even) nucleons, these shells  separate into three categories in sHF-EFA
\begin{enumerate}
\item $\epsilon_{\breve{h}}$ denoting ``hole states",
\item $\epsilon_{\breve{v}}$ denoting ``valence states",
\item $\epsilon_{\breve{p}}$ denoting ``particle states" ,
\end{enumerate}
such that $A-a_v$ nucleons fill the hole states whereas $0<a_v\leq d_v$ nucleons occupy the valence shell.

\begin{figure}
    \centering
    \includegraphics[width=0.5\textwidth]{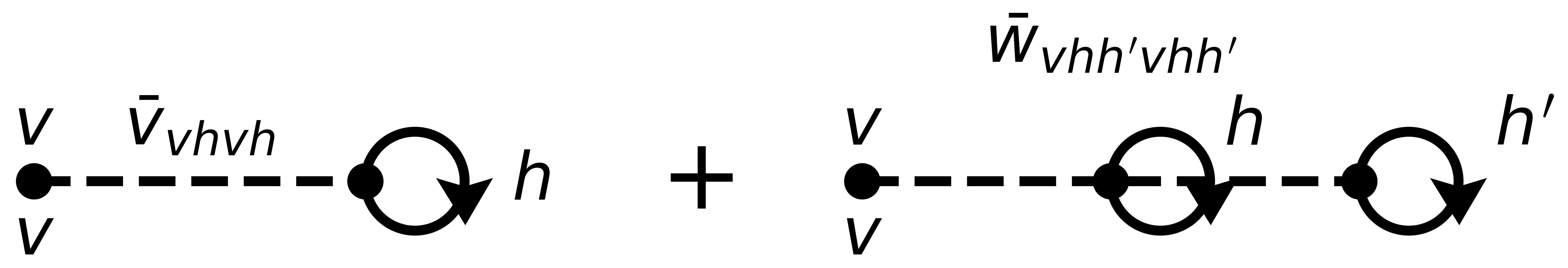}
    \caption{First-order interaction energy contributions to the valence-shell single-particle energy $\epsilon^{\text{CS}}_{\breve{v}}$ computed in the closed-shell core.}
    \label{diagrams1storder_e}
\end{figure}

\begin{figure}
    \centering
    \includegraphics[width=0.5\textwidth]{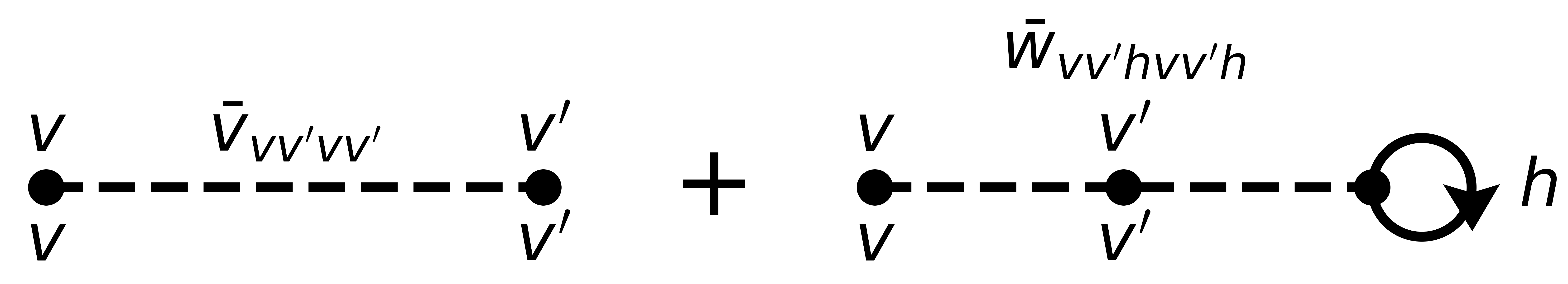}
    \caption{First-order contributions to the valence-shell effective two-body matrix elements $\overline{\bold{v}}_{vv'vv'}$.}
    \label{diagrams1storder_v}
\end{figure}

Given this setting, one eventually obtains the total energy of an open-shell nucleus relative to the CS core along with the corresponding two-neutron separation energy and two-neutron shell gap as
\begin{subequations}
\label{EFA}
\begin{align} 
\Delta E^\text{sHF-EFA}(a_{v}) &\equiv  E^\text{sHF-EFA}(a_{v}) -  E^\text{sHF-EFA}(0)\nonumber \\ 
&= \alpha_{\breve{v}}  a_v + \frac{\beta_{\breve{v}}}{2} a^2_{v} \, ,  \label{EFA1} \\
S^\text{sHF-EFA}_{2n}(a_{v})&= -2\alpha_{\breve{v}} -2\beta_{\breve{v}}( a_{v} -1) \, , \label{EFA2} \\
\Delta^\text{sHF-EFA}_{2n}(a_{v}) &= 4 \beta_{\breve{v}}  \, . \label{EFA3}
\end{align}
\end{subequations}
with 
\begin{subequations}
\label{EFAderivativesbulk} 
\begin{align}
\alpha_{\breve{v}} &  = \epsilon^{\text{CS}}_{\breve{v}}  \nonumber \\
&\equiv t_{vv} + \sum_{h} \overline{v}_{vhvh} + \frac{1}{2} \sum_{hh'} \overline{w}_{vhh'vhh'}\, , \label{EFAderivativesbulk1} \\
\beta_{\breve{v}} & = \frac{1}{d_{v}} \sum_{m_{v'}}^{d_{v}} \left(\overline{v}_{vv'vv'} + \sum_{h} \overline{w}_{vv'hvv'h}\right)  \nonumber \\
&\equiv \frac{1}{d_{v}} \sum_{m_{v'}}^{d_{v}} \overline{\bold{v}}_{vv'vv'} \label{EFAderivativesbulk2}  \, .
\end{align}
\end{subequations}
Equation~\eqref{EFA1} proves that the sHF-EFA energy is indeed quadratic\footnote{As shown in~\ref{analytical}, the 3N interaction actually induces the presence of a cubic term in the energy. However, present numerical applications demonstrate that it is negligible for all nuclei under consideration such that it can be dropped altogether in the present discussion.} in the number of valence nucleons throughout any given open-shell. The coefficient $\alpha_{\breve{v}}$ of the linear term is nothing but the mean-field single-particle energy of the valence shell computed in the CS core $\epsilon^{\text{CS}}_{\breve{v}}$, whose interaction energy contributions are displayed diagrammatically in Fig.~\ref{diagrams1storder_e}. The coefficient $\beta_{\breve{v}}$ of the quadratic term, i.e. the curvature of the energy, is given by the {\it average} over the valence magnetic substates of the diagonal valence-shell two-body matrix elements\footnote{As seen from Eq.~\eqref{EFAderivativesbulk2}, $\overline{\bold{v}}_{vv'vv'}$ includes the effective contribution obtained by averaging the 3N interaction over the CS core.} $\overline{\bold{v}}_{vv'vv'}$ displayed diagrammatically in Fig.~\ref{diagrams1storder_v}. Such an averaging corresponds to the {\it monopole} valence-shell matrix element per valence state. As visible from Eq.~\eqref{EFA2}, $-2\epsilon^{\text{CS}}_{\breve{v}}$ sets the initial value of $S_{2n}$\footnote{As seen from Tab.~\ref{fig:coefficients}, the relation $|\alpha_{\breve{v}}|\gg|\beta_{\breve{v}}|$ holds in practice such that the starting value of $S^\text{sHF-EFA}_{2n}$ ($a_v=2$) in the open shell is essentially dictated by $\epsilon^{\text{CS}}_{\breve{v}}$.} whereas $-2\beta_{\breve{v}}$ drives its linear evolution throughout the open-shell. Eventually, $\Delta_{2n}$ extracts $4\beta_{\breve{v}}$.

\begin{figure}
    \centering
    \includegraphics[width=0.5\textwidth]{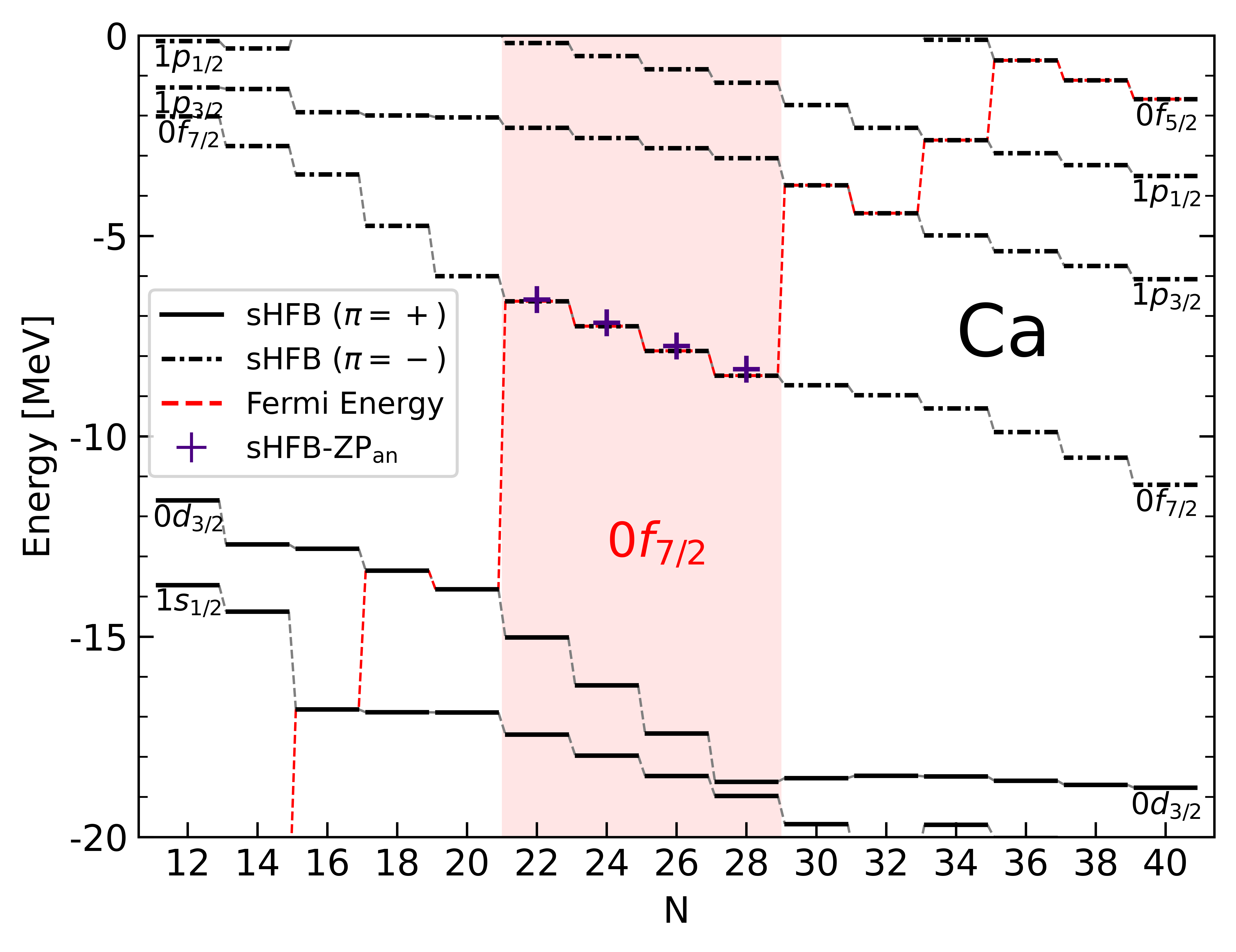}
    \caption{Neutron canonical single-particle energies $\epsilon_{\breve{k}}$ from sHFB calculations along the Ca isotopic chain. Semi-analytical sHFB-ZP results for $\epsilon_{0\text{f}_{7/2}}$ are also shown between $^{40}$Ca and $^{48}$Ca.}
    \label{sHFBespe}
\end{figure}

Extracting $\epsilon^{\text{CS}}_{\breve{v}}$ and $\overline{\bold{v}}_{vv'vv'}$ numerically from the presently employed chiral Hamiltonian (see Tab.~\ref{fig:coefficients}), the semi-analytical results from Eqs.~\eqref{EFA}-\eqref{EFAderivativesbulk} (in fact of their sHFB-ZP counterparts; see~\ref{analytical}) are superimposed on the left panels of Fig.~\ref{sHFB} between $^{42}$Ca and $^{48}$Ca (0f$_{7/2}$ valence shell) as well as between $^{62}$Ca and $^{70}$Ca (0g$_{9/2}$ valence shell). The results perfectly match the numerical sHF-EFA curves, that are themselves very close to sHFB results. Looking at the left panels of Fig.~\ref{sHFBcurvature}, one indeed sees the fully quantitative agreement between sHF-EFA and the semi-analytical results.

\begin{table}
\centering
\begin{tabular}{l|c|c}
\hline
Open shell & $\alpha_{\breve{v}}$ (MeV) & $\beta_{\breve{v}}$ (MeV) \\
\hline
0f$_{7/2}$ & -6.005	& -0.290 \\
\hline
0g$_{9/2}$ & 2.976 & -0.270 \\
\hline
\end{tabular}
\caption{Coefficient of the linear and quadratic term of the HFB-ZP energy (Eq.~\eqref{HFBZP}-\eqref{derivatives}) extracted numerically for two neutron valence shells along the Ca isotopic chain using the EM\,1.8/2.0 Hamiltonian~\cite{Hebe11fits}. The coefficient of the cubic term $\gamma_{\breve{v}}$ is numerically zero in all cases.}
\label{fig:coefficients}
\end{table}

The semi-analytical results first clarify that, in an \ai{} setting, the reason why in a given open shell 
\begin{enumerate}
\item $E^\text{sHFB}$ loses energy relatively to experiment,
\item $S^\text{sHFB}_{2n}$ starts from too low a value,
\end{enumerate}
relates directly to the fact that the mean-field valence-shell single-particle energy in the CS core $\epsilon^{\text{CS}}_{\breve{v}}$ delivered by $\chi$EFT interactions is systematically too small in absolute, i.e. non negative enough. This is accompanied with the fact that the effective mass is too low at the mean-field level, as testified by the too large (value) decrease of  $S^\text{sHFB}_{2n}$ ($\Delta^\text{sHFB}_{2n}$), which is actually a key reason why pairing correlations are so weak. Second, the fact that
\begin{enumerate}
\item $E^\text{sHFB}$ is concave,
\item $S^\text{sHFB}_{2n}$ is rising,
\item $\Delta^\text{sHFB}_{2n}$ is negative,
\end{enumerate}
throughout open shells, in opposition to experimental data, relates to the attractive character of the monopole valence-shell matrix element delivered by $\chi$EFT interactions. 

Interestingly, the above features are typically {\it not} displayed by sHFB calculations based on  {\it effective} and empirical energy density functionals (EDF), see e.g.~\cite{Burrello:2021nsf}. Indeed, EDFs are tailored via a fit to empirical data to implicitly incorporate the dominant effect of dynamical correlations. In practice, this generally results into a significantly larger effective mass\footnote{Even though $\Delta_{2n}$ is traditionally left to overestimate experimental data at shell closures in EDF calculations to leave some room for additional correlations, it does so on a much smaller scale than in present sHFB \ai{} calculations that overestimate $\Delta_{2n}$  at, e.g. $N=20$ by $4$\, MeV.} and into much stronger pairing correlations since the pairing part of the functional is typically adjusted to reproduce experimental pairing gaps at the sHFB level. At the same time, it is striking that EDF parametrizations only tailored to reproduce many-body calculations of infinite nuclear matter and employed in finite nuclei at the strict mean-field level, i.e. without an explicit account of dynamical correlations on a nucleus-by-nucleus basis, do display the features identified above~\cite{Burrello:2021nsf}.

\begin{figure*}
    \centering
    \includegraphics[width=1.0\textwidth]{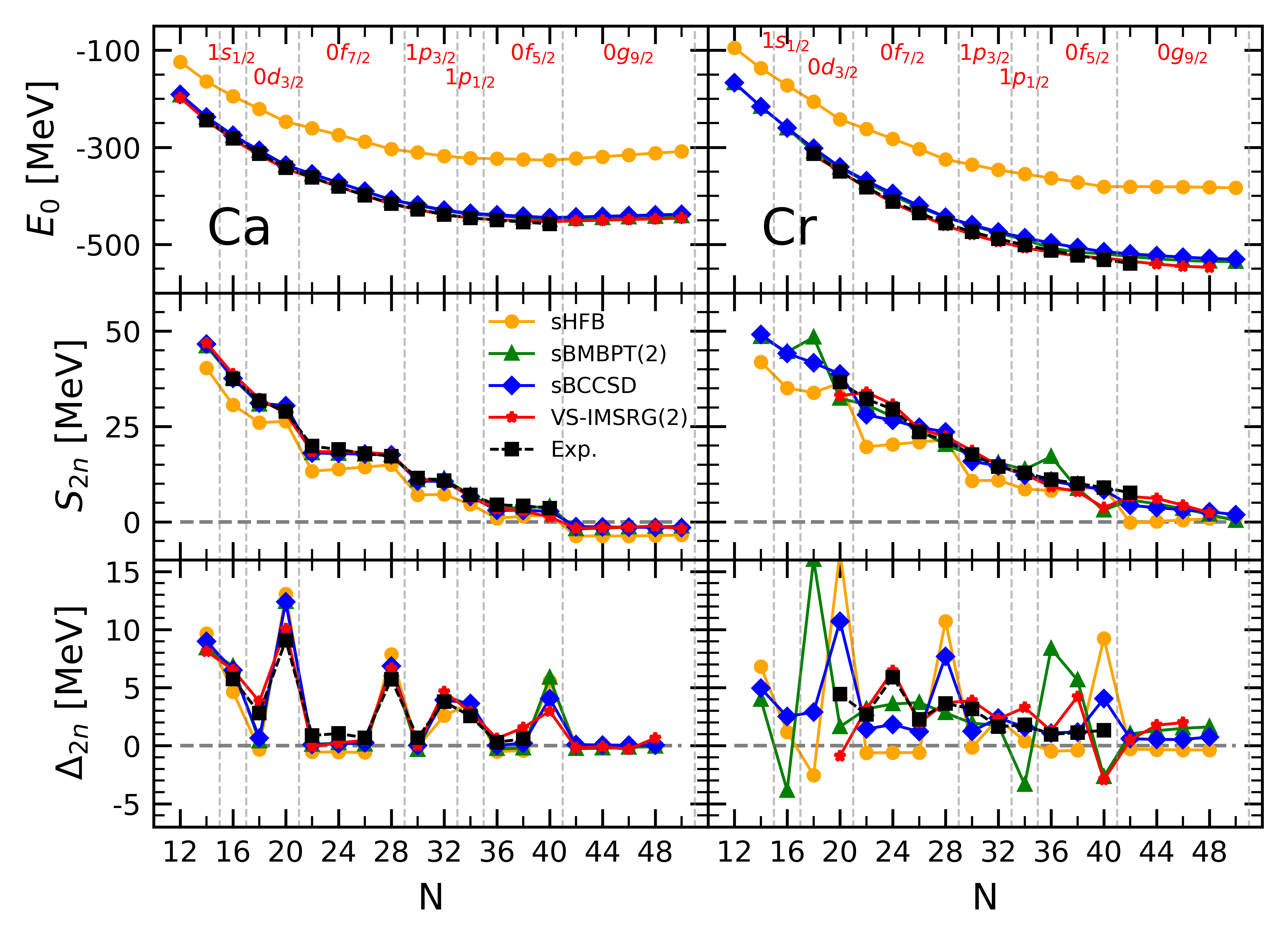}
    \caption{Results of sHFB, sBMBPT(2), sBCCSD and VS-IMSRG(2) calculations against experimental data along Ca (left) and Cr (right) isotopic chains. Upper panels: absolute binding energies. Middle panels: two-neutron separation energy. Lower panels: two-neutron shell gap.}
    \label{sbeyondHFB}
\end{figure*}

The above observations are also consistent with the evolution of canonical single-particle energies throughout an open shell, and more specifically of the  valence-shell single-particle energy itself. In the sHF-EFA approximation, it can easily be shown that its evolution with $a_{v}$ is linear
\begin{align}
\epsilon^\text{sHF-EFA}_{\breve{v}}(a_{v}) =& \epsilon^{\text{CS}}_{\breve{v}} + \beta_{\tilde{v}} a_{v} \, , \label{speEFAbulk}
\end{align}
the coefficient of the slope being given by $\beta_{\tilde{v}}$. As visible in Fig.~\ref{sHFBespe}, neutron canonical single-particle energies do evolve linearly within a given open-shell. In particular, the evolution of $\epsilon_{0\text{f}_{7/2}}$ between $^{40}$Ca and  $^{48}$Ca is perfectly reproduced using Eq.~\eqref{speEFAbulk}  (in fact its sHFB-ZP counterpart; see \ref{analytical}). Eventually, this linear down-slopping evolution is fully correlated with the concavity of the binding energy.

\subsection{Cr chain}
\label{sMeanfieldsyst2}

Having characterized sHFB results along the semi-magic Ca isotopic chain, we focus on to doubly open-shell Cr isotopes. 

As seen in the upper-right panel of Fig.~\ref{sHFB}, the global trend of sHFB binding energies is similar, relative to the data, than for Ca isotopes.  In the magnifying glass of $S_{2n}$ and  $\Delta_{2n}$, experimental data do not however display the characteristic patterns identified along the Ca chain. In particular, $S_{2n}$ decreases more gradually such that the sudden drops (sudden spikes in $\Delta_{2n}$), e.g. at $N=20$ and $28$, have all disappeared. Contrarily, a small bump (spike) is now visible in $S_{2n}$ ($\Delta_{2n}$) for $N=24$, i.e. in $^{44}$Cr located in the middle of the 0f$_{7/2}$ shell. These changes are not at all accounted for by sHFB results that closely follow those obtained previously. Indeed, in addition to displaying the defects identified along the Ca isotopic chain, sHFB results further fail to capture the qualitative modifications seen in the data, i.e. sHFB results keep a strong memory of the underlying spherical shell structure whose fingerprints are no longer visible along the Cr isotopic chain. 

\section{Spherical beyond mean-field corrections}
\label{sBeyonMeanfield}

Based on the previous analysis, the goal is now to assess whether consistently adding dynamical correlations via sBMBPT(2), sBCCSD or VS-IMSRG(2) can correct for the shortcomings identified at the sHFB level.

\begin{figure}
    \centering
    \includegraphics[width=0.5\textwidth]{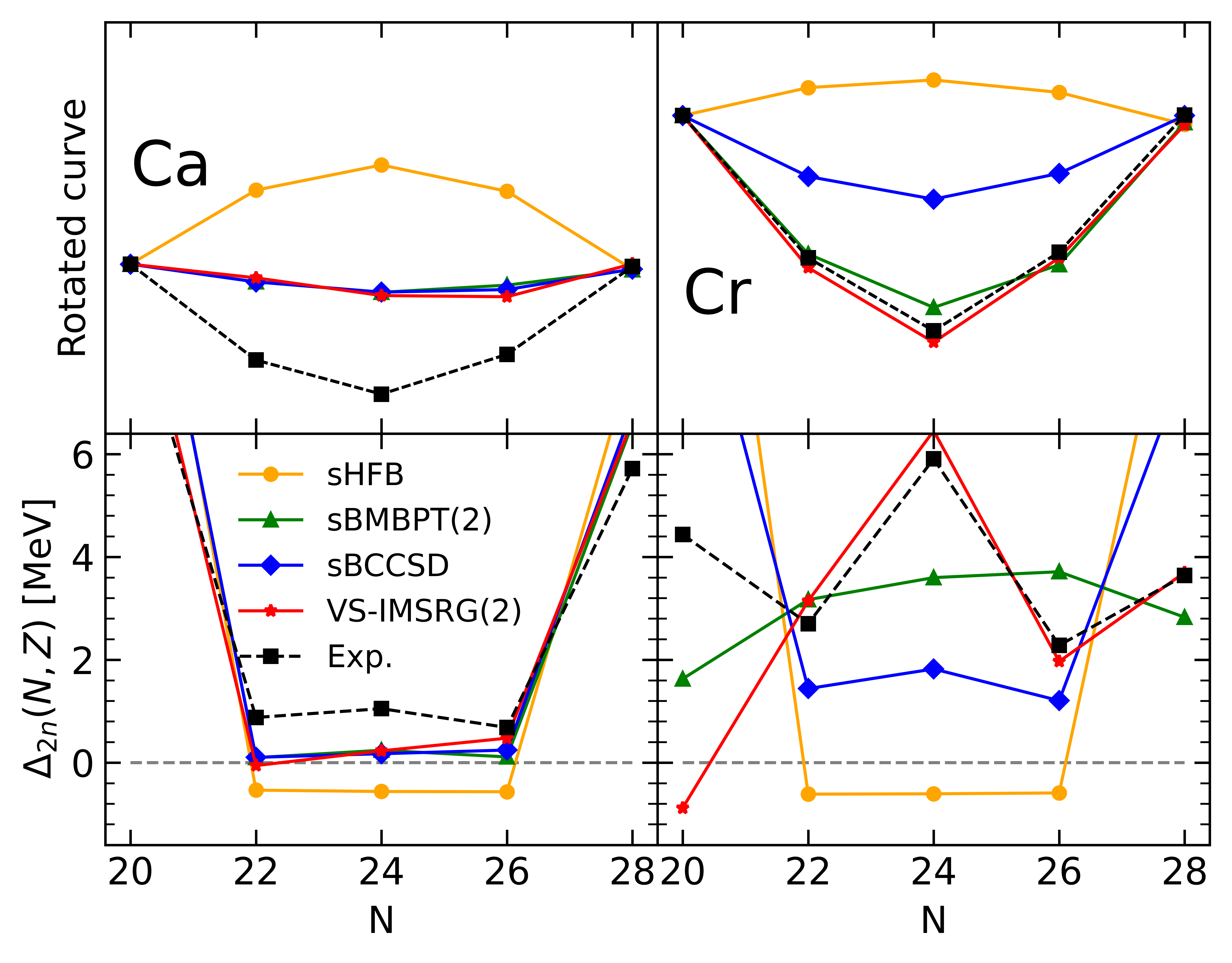}
    \caption{Same as Fig.~\ref{sHFBcurvature} for sHFB, sBMBPT(2), sBCCSD and VS-IMSRG(2).}
    \label{sbeyondHFBcurvature}
\end{figure}

\subsection{Ca chain}
\label{sbeyondMeanfieldsyst1}

As seen in the upper-left panel of Fig.~\ref{sbeyondHFB}, dynamical correlations compensate for the underbinding observed at the sHFB level such that all three methods reproduce well experimental binding energies along the Ca isotopic chain with the presently employed Hamiltonian. This is particularly true for VS-IMSRG(2) whose root-mean-square error to the data is equal to $1.9$\,MeV, while it is equal to $7.3$ and $8.8$\,MeV for BMBPT(2) and BCCSD, respectively. In particular, the increasing underbinding of sHFB results as a function of neutron excess is essentially compensated for. 

The improvement goes indeed beyond a plain shift as can be inferred from the middle-left panel  of Fig.~\ref{sbeyondHFB}. Indeed, $S_{2n}$ are systematically improved against experimental data for all three methods. First, $S_{2n}$ are globally increased by up to about $5$\,MeV. Second, the amplitudes of the sudden drops at magic numbers are reduced. As visible from the bottom-left panel, the two-neutron shell gap at $N=20$ is reduced from $13$\,MeV in sHFB to $8.6$\,MeV in VS-IMSRG(2), which is comparable to the experimental value of $9.1$\,MeV. In sBMBPT(2) and sBCCSD the reduction is not pronounced enough, the $\Delta_{2n}$ being equal to $12.4$ in both cases, thus showing that low-rank elementary excitations are not enough to produce a fully quantitative picture of the $N=20$ magicity. While sBCCSD is third-order-complete, it is of interest to investigate how much including genuine fourth-order triple excitations, e.g. by going to (approximate) BCCSDT, can help in this respect~\cite{vernik24a}. 

In spite of the $N=20$ two-neutron shell gap being still overestimated in  sBMBPT(2) and sBCCSD, the $S_{2n}$ at the beginning of each open shell is increased to be in much better agreement with experimental data. For example, dynamical correlations bring $S_{2n}$ in $^{42}$Ca from $13.3$\,MeV in sHFB to $18.0$ and $18.1$\,MeV in BMBPT(2) and BCCSD, respectively, as well as to $18.6$\,MeV in VS-IMSRG(2), which compares favorably with the experimental value of $19.8$\,MeV. Third, the wrong linear increase throughout any given open shell is corrected for, as can be seen for example between $^{42}$Ca and $^{48}$Ca. This reflects the improvement of the curvature of the energy throughout open shells that can be better appreciated from the left-panels of Fig.~\ref{sbeyondHFBcurvature} that focuses on the 0f$_{7/2}$ shell. Dynamical correlations turn the energy from beyond concave at the sHFB level to being convex, in a way that is essentially identical with the three employed methods. 

Eventually, the agreement with data for $S_{2n}$ and $\Delta_{2n}$ along the Ca chain is qualitatively and quantitatively satisfying for all three methods even though the $N=20$ magicity is still exaggerated in BMBPT(2) and BCCSD and the convexity throughout the 0f$_{7/2}$ shell is not pronounced enough compared to experimental data for all three methods, which points to yet missing correlations. It will be interesting to investigate in the future whether the lack of convexity in the energy is correlated with the inability of presently employed \ai{} methods to correctly reproduce the (infamous) evolution of charge radii between $^{40}$Ca and $^{48}$Ca~\cite{Caurier:2001np}.

\begin{figure}
    \centering\includegraphics[width=0.5\textwidth]{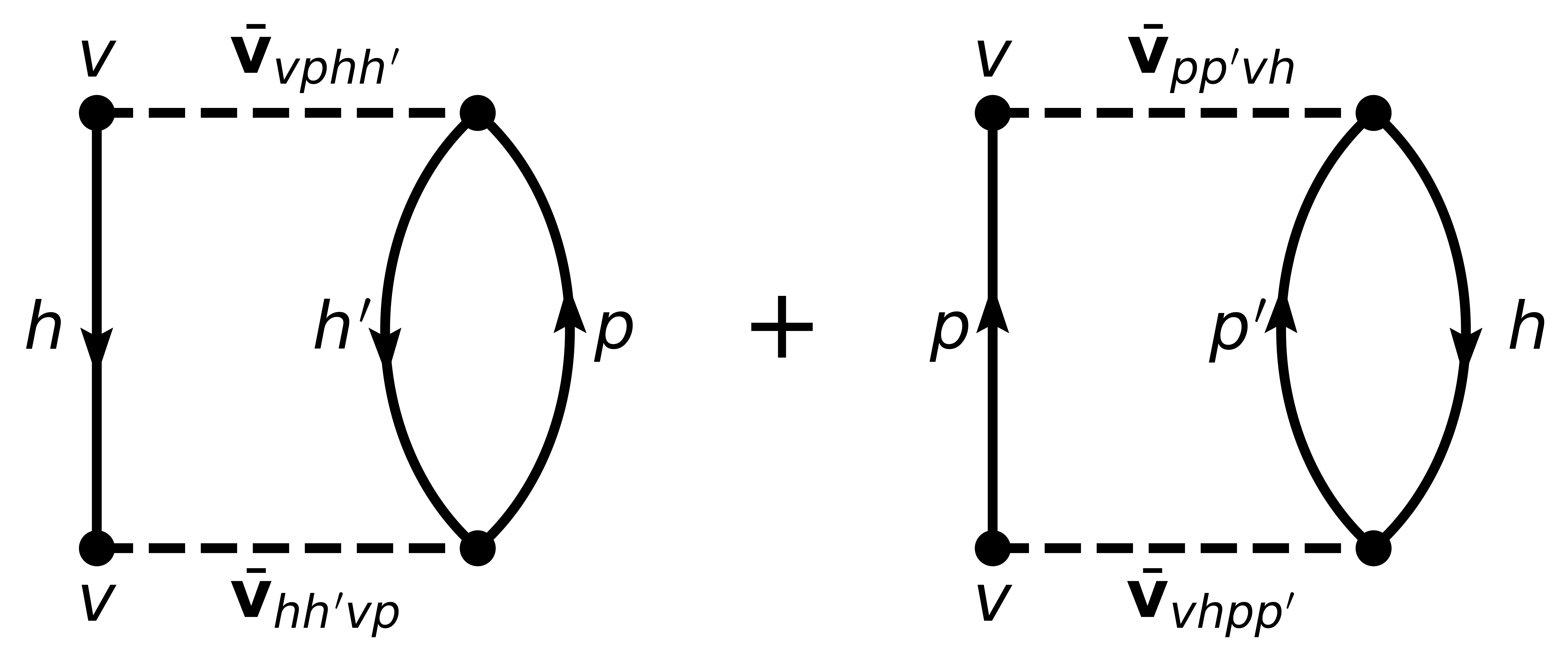}
    \caption{Second-order (on-shell) diagonal self-energy correction $\Sigma^{(2)}_{\breve{v}}$ to the valence-shell single-particle energy computed in the closed-shell core. Left: 1-particle/2-hole diagram. Right: 2-particle/1-hole diagram.}
    \label{diagrams2ndorder_e}
\end{figure}

\begin{figure}
    \centering\includegraphics[width=0.5\textwidth]{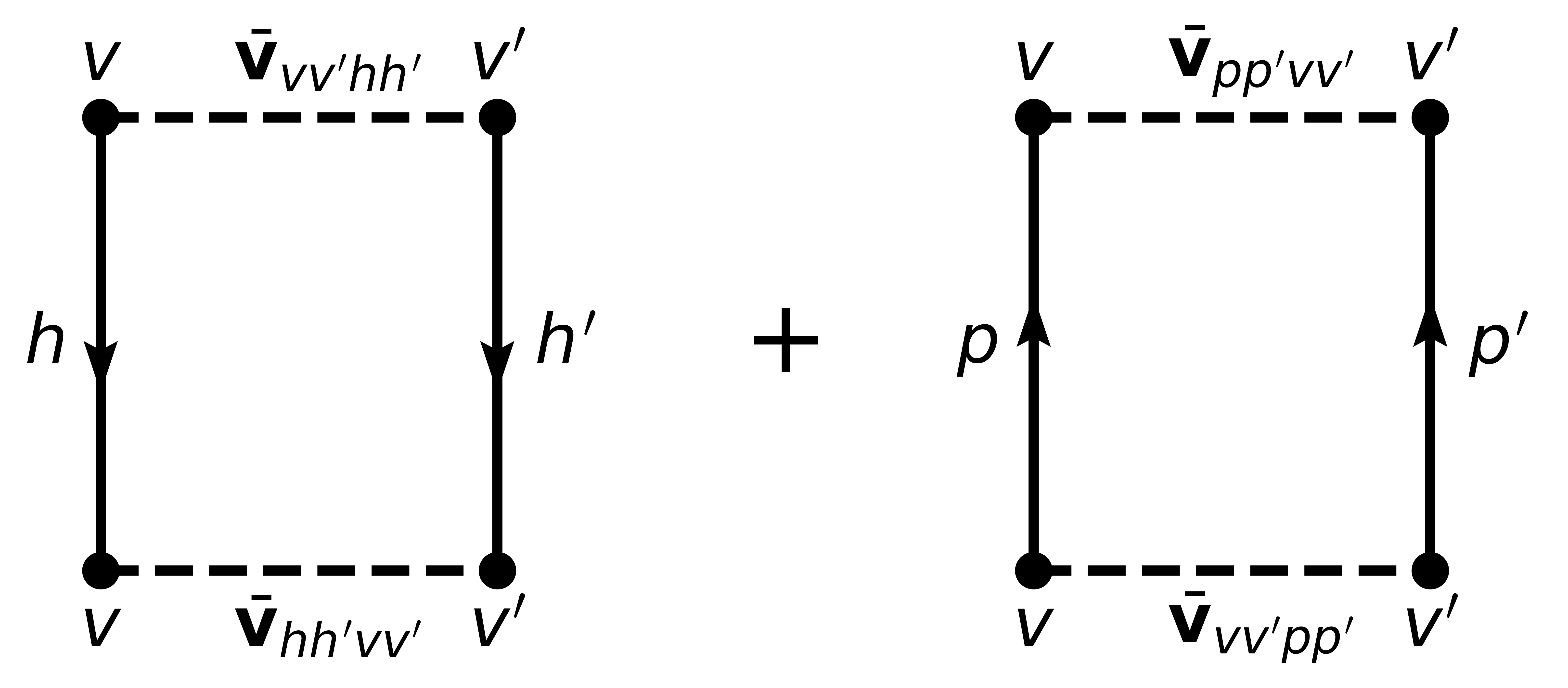}
    \caption{Second-order (on-shell) correction $\bar{v}^{(2)}_{vv'vv'}$ to the diagonal valence-shell effective two-body matrix elements. Left: hole-hole diagram. Right: particle-particle diagram.}
    \label{diagrams2ndorder_v}
\end{figure}

\subsection{Analytical investigation}
\label{sbeyondMeanfieldanalytic}

As demonstrated in Sec.~\ref{sMeanfieldanalytic}, the deficiencies of sHFB can be understood via a semi-analytical analysis performed in the zero-pairing limit. The capacity of dynamical correlations to correct for those shortcomings is now analyzed in a similar manner within the frame of sMBPT(2).  As demonstrated in~\ref{MBPT2app}, the mean-field result of Eq.~\eqref{EFA} can be extended, for $a_v \geq 2$, to
\begin{subequations}
\label{EFA2order}
\begin{align} 
S^\text{(2)}_{2n}(a_{v})&= -2\epsilon^{\text{CS}(2)}_{\breve{v}}   - 2\beta^{(2)}_{\breve{v}} (a_{v} -1)  \, , \label{EFA2order1} \\
\Delta^\text{(2)}_{2n}(a_{v}) &= 4 \beta^{(2)}_{\breve{v}}  \, , \label{EFA2order2}
\end{align}
\end{subequations}
where the second-order (on-shell) valence-shell single-particle energy and averaged valence-shell interaction computed in the CS core
\begin{subequations}
\label{2ordercorr}
\begin{align} 
\epsilon^{\text{CS}(2)}_{\breve{v}} &\equiv \epsilon^{\text{CS}}_{\breve{v}} + \Sigma^{(2)}_{\breve{v}}(\epsilon^{\text{CS}}_{\breve{v}}) \, , \label{2ordercorr_e} \\
\beta^{(2)}_{\tilde{v}} &\equiv \frac{1}{d_{v}} \sum_{m_{v'}}^{d_{\breve{v}}} \left(\overline{\bold{v}}_{vv'vv'} + \overline{\bold{v}}^{(2)}_{vv'vv'}(\epsilon^{\text{CS}}_{\breve{v}})\right) \, . \label{2ordercorr_v}
\end{align}
\end{subequations}
involve the (on-shell) valence-shell self-energy and two-body effective interaction corrections
\begin{subequations}
\label{corrections2ndorderbulk}
\begin{align} 
\Sigma^{(2)}_{\breve{v}}(\epsilon^{\text{CS}}_{\breve{v}}) &= +\frac{1}{2} \sum_{hh'p} \frac{|\overline{\bold{v}}_{hh'vp}|^2}{\epsilon^{\text{CS}}_{p}+\epsilon^{\text{CS}}_{\breve{v}}-\epsilon^{\text{CS}}_{h}-\epsilon^{\text{CS}}_{h'}} \nonumber \\
&\hspace{0.4cm} - \frac{1}{2} \sum_{pp'h} \frac{|\overline{\bold{v}}_{vhpp'}|^2}{\epsilon^{\text{CS}}_{p}+\epsilon^{\text{CS}}_{p'}-\epsilon^{\text{CS}}_{h}-\epsilon^{\text{CS}}_{\breve{v}}} \label{corrections2ndorderbulk1}  \\
\overline{\bold{v}}^{(2)}_{vv'vv'}(\epsilon^{\text{CS}}_{\breve{v}}) &=  +\frac{1}{2} \sum_{hh'}  \frac{|\overline{\bold{v}}_{hh'vv'}|^2}{2\epsilon^{\text{CS}}_{\breve{v}}-\epsilon^{\text{CS}}_{h}-\epsilon^{\text{CS}}_{h'}} \nonumber \\
&\hspace{0.4cm} - \frac{1}{2} \sum_{pp'} \frac{|\overline{\bold{v}}_{vv'pp'}|^2}{\epsilon^{\text{CS}}_{p}+\epsilon^{\text{CS}}_{p'}-2\epsilon^{\text{CS}}_{\breve{v}}} \label{corrections2ndorderbulk2} \, ,
\end{align}
\end{subequations}
displayed diagrammatically in Figs.~\ref{diagrams2ndorder_e} and~\ref{diagrams2ndorder_v}, respectively. The self-energy correction collects a positive (2-hole/1-particle) contribution and a negative (1-hole/2-particle) contribution. Similarly, the valence-shell interaction correction  collects a positive (hole-hole) contribution and a negative (particle-particle) contribution. 

As seen from Eqs.~\eqref{EFA2order}-\eqref{corrections2ndorderbulk}, and in agreement with the results shown in the middle-left panel of Fig.~\ref{sbeyondHFB} and analyzed in the present section, dynamical correlations modify both the starting value and the slope of $S_{2n}$ in the valence-shell. For example, the negative second-order self-energy correction $\Sigma^{(2)}_{0\text{f}_{7/2}}$ lowers $\epsilon^{\text{CS}(2)}_{0\text{f}_{7/2}}$ in such a way that $S_{2n}$ computed in sBMBPT(2) increases from $13.28$ to $18.02$\,MeV in $^{42}$Ca to almost match the experimental value ($19.84$\,MeV). This effect relates to the coupling of a propagating nucleon to 1-particle/2-hole and 2-particle/1-hole configurations as represented in Fig~\ref{diagrams2ndorder_e}, the latter winning over the former\footnote{The lowering of $\epsilon_{0\text{f}_{7/2}}$ is not accompanied by a decrease of $\Delta_{2n}$ in $^{40}$Ca in sBMBPT(2) and sBCCSD, contrary to sVS-IMSRG(2), i.e. $\epsilon_{0\text{d}_{3/2}}$ is lowered as much as $\epsilon_{0\text{f}_{7/2}}$. Thus, the needed increase of the effective mass associated with the compression of on-shell single-particle energies is not accounted for by low-order corrections to sHFB.}. Consistently, the second-order correction to the average 0f$_{7/2}$ valence-shell effective interaction is repulsive, with the hole-hole contribution winning over the particle-particle one. In the present calculation, such a correction is larger in absolute value than the mean-field contribution and manages to turn the total energy from being concave to being convex, i.e. it makes $S_{2n}$ decrease linearly between $^{42}$Ca and $^{48}$Ca as for experimental data\footnote{The amount by which $S_{2n}$ is increased at the start of the open-shell and the fact that its slope is actually inverted depend on the Hamiltonian under use; see Refs.~\cite{Tichai18BMBPT,Soma:2020dyc} for examples where the qualitative defects of the sHFB results are not actually corrected via the inclusion of low-order dynamical correlations.}. Still, and as can be seen from the bottom-left panel of Fig.~\ref{sbeyondHFBcurvature}, the  positive curvature $\beta^{(2)}_{0\text{f}_{7/2}}=25$\,keV\footnote{This value is essentially constant throughout the valence shell.} is not large enough\footnote{The same is true for sBCCSD and VS-IMSRG(2) calculations as can be inferred from the bottom-left panel of Fig.~\ref{sbeyondHFBcurvature}.} compared to experimental data ($\Delta_{2n}/4\approx 220$\,keV in $^{42-46}$Ca), thus pointing to yet missing many-body correlations as discussed earlier on.

\subsection{Cr chain}
\label{sbeyondMeanfieldsyst2}

While the deficiencies observed at the sHFB were shown to be qualitatively and quantitatively corrected via the consistent addition of dynamical correlations in Ca isotopes, it remains to be seen to which extent this is the case along the Cr isotopic chain.

As seen in the upper-right panel of Fig.~\ref{sbeyondHFB}, correlations brought by sBMBPT(2), sBCCSD and VS-IMSRG(2) provide the bulk of the missing binding along the Cr chain as well, even though the end values are globally further away from experimental data than for Ca isotopes. While the rms error to the data is $1.9$, $7.3$ and $8.8$\,MeV for VS-IMSRG(2), sBMBPT(2) and sBCCSD in Ca isotopes, it becomes $4.0$, $10.6$ and $14.7$\,MeV in Cr isotopes, respectively; i.e. the deterioration is more pronounced for sBMBPT(2) and sBCCSD. 

Looking at the middle- and bottom-right panels of Fig.~\ref{sbeyondHFB}, sBMBPT(2) and sBCCSD are seen to improve the reproduction of experimental $S_{2n}$ and $\Delta_{2n}$ compared to sHFB. Still, the level of agreement is neither on the same level as in Ca isotopes nor on the same level as for VS-IMSRG(2) in those Cr isotopes. The large spikes of $\Delta_{2n}$ seen at $N=20, 28$ and $40$ for sHFB are only slightly diminished in sBCCSD calculations, thus wrongly keeping the imprint of the spherical magic numbers. Even if the behavior throughout the 0f$_{7/2}$ shell is improved, as can also be appreciated from the left panels of Fig.~\ref{sbeyondHFBcurvature}, it remains quite remote from experimental data. As for  sBMBPT(2) results, $\Delta_{2n}$ bear little resemblance to experimental data and are clearly not credible. 

Contrarily, the $S_{2n}$ and $\Delta_{2n}$ predicted by VS-IMSRG(2) are both in qualitative and quantitative agreement with experimental data\footnote{The slight degradation observed in the vicinity on $N=20$ and $40$ is attributable to the need to reset the valence space.}. Indeed, the disappearance of the spikes at $N=20, 28$ and $40$, as well as the appearance of a new one for $N=24$, are perfectly reproduced. This demonstrates that the exact diagonalization of the effective Hamiltonian within the $fp$ shell is able to capture crucial static correlations that are not accounted for by low-rank excitations on top of a spherical mean field via sBMBPT(2) and sBCCSD.

\section{Deformed unperturbed state}
\label{dMeanfieldsyst} 

Even if challenges remain to be overcome to reach high accuracy or the description of specific observables impacted by collective fluctuations (e.g. superfluidity, radii between $^{40}$Ca and $^{48}$Ca\ldots), the discussion above demonstrates that polynomially-scaling expansion methods built on top of a spherical Bogoliubov reference state and implemented to rather low truncation order deliver a good account of mid-mass doubly closed-shell and singly open-shell nuclear ground states. Contrarily, doubly open-shell nuclei require the inclusion of specific static correlations that can hardly be incorporated following this strategy, i.e. they require a full diagonalization of the effective Hamiltonian in an appropriate valence space, thus compromising with the polynomial scaling that will eventually become crucial in heavy nuclei. 

\begin{figure*}
    \centering
    \includegraphics[width=1.0\textwidth]{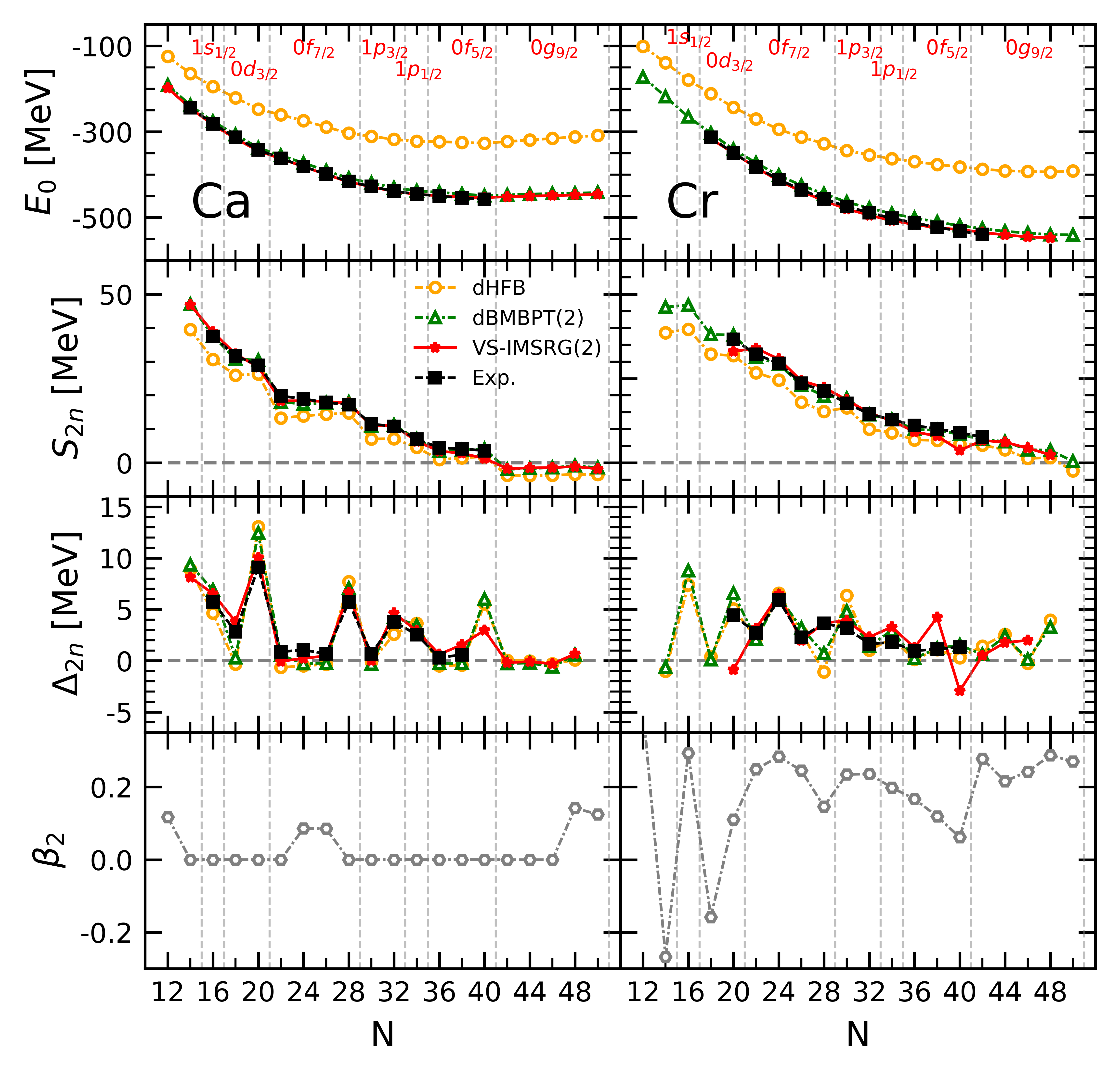}
    \caption{Systematic dHFB, dBMBPT(2) and VS-IMSRG(2) calculations against experimental data along Ca (left) and Cr (right) isotopic chains. First line: absolute binding energies. Second line: two-neutron separation energy. Third line: two-neutron shell gap. Fourth line: intrinsic axial quadrupole deformation of the dHFB solution.}
    \label{dbeyondHFB}
\end{figure*}

\begin{figure}
    \centering
    \includegraphics[width=0.5\textwidth]{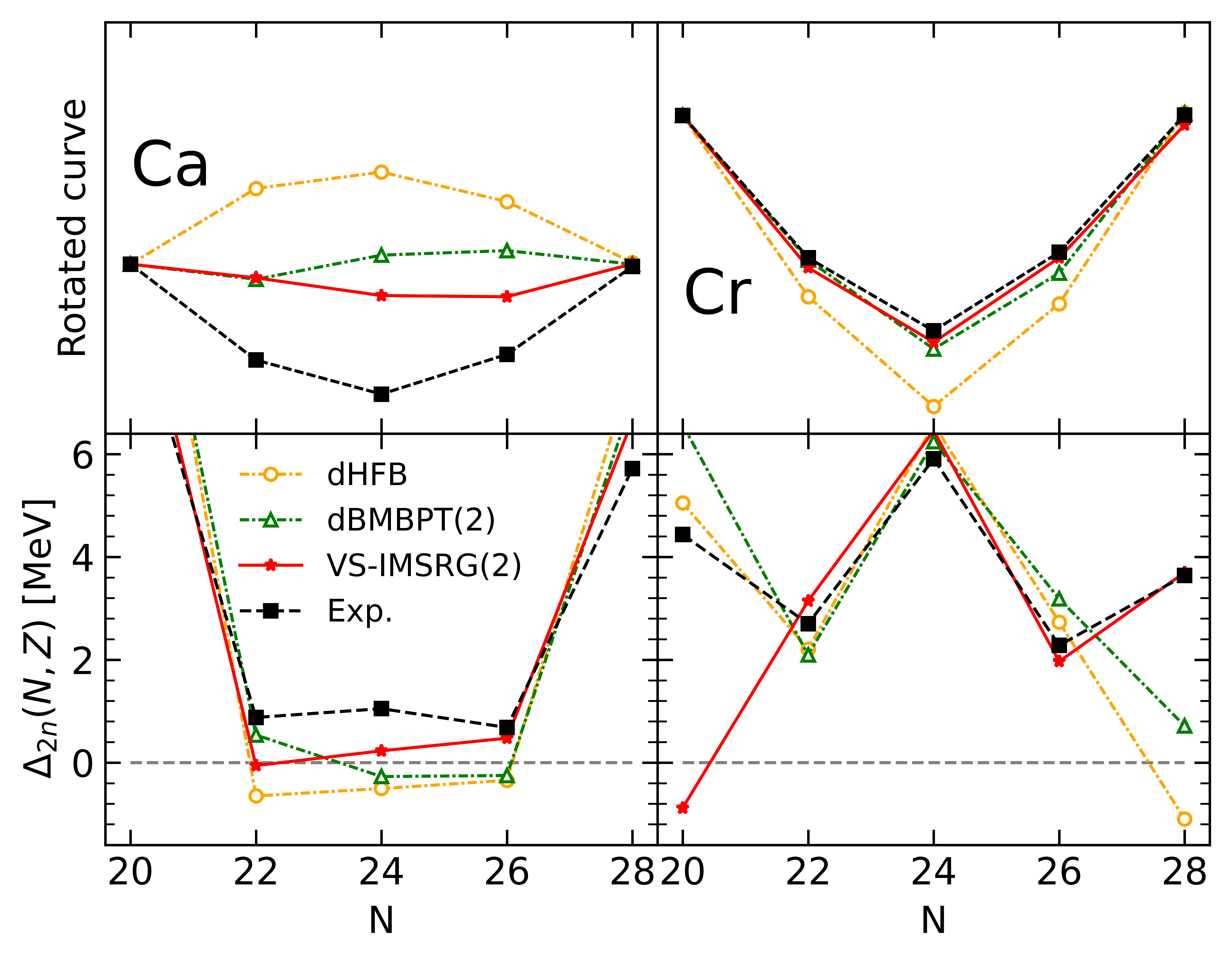}
    \caption{Same as Fig.~\ref{sHFBcurvature} for dHFB, dBMBPT(2) and VS-IMSRG(2).}
    \label{dbeyondHFBcurvature}
\end{figure}

On a principle level, the solution delivered by expansion many-body methods is eventually independent of the unperturbed state whenever all terms in the expansion series are summed up -- provided that the expansion series actually converges~\cite{leininger00a,RothPade,Langhammer:2012jx,Tichai:2016joa,Demol:2020mzd,Demol:2020nm}. In practice however, the interesting question relates to how close to the exact solution one can be at the most economical cost. In this context, it is believed that dominant static correlations can be efficiently captured in doubly open-shell nuclei via an appropriate redefinition of the unperturbed state, at the price of breaking~\cite{Novario:2020kuf,Frosini:2024ajq} (and eventually restoring~\cite{Duguet:2014jja,Yao:2019rck,Hagen:2022tqp,Frosini:2021ddm,Sun:2024iht}) rotational symmetry associated with angular-momentum conservation. The present section wishes to illustrate that a quantitative description of doubly open-shell nuclei can indeed be achieved at (low) polynomial cost via dBMBPT(2) calculations performed on top of a deformed HFB unperturbed state.

\subsection{Ca chain}
\label{dMeanfieldsyst1}

Results of systematic dHFB, dBMBPT(2), as well as VS-IMSRG(2) calculations of Ca isotopes are displayed on the left-hand panels of Fig.~\ref{dbeyondHFB}. Comparing those to the results shown before on the left-hand panels of Fig.~\ref{sbeyondHFB}, it is clear that allowing the mean-field solution to deform does not lead to any significant modification along the Ca isotopic chain. Indeed, and as demonstrated by the lower panel of Fig.~\ref{dbeyondHFB}, almost all Ca isotopes do not take advantage of this possibility at the mean-field level\footnote{The few isotopes that do deform, i.e. $^{32,44,46,68,70}$Ca, only acquire a small intrinsic deformation.}. The fact that static correlations associated with quadrupolar deformations are not emerging from the calculation is consistent with the fact sBMBPT(2) and sBCCSD results were already satisfactory as discussed extensively in Sec.~\ref{sBeyonMeanfield}.

\subsection{Cr chain}
\label{dMeanfieldsyst2}

As the comparison of the right-hand panels of Figs.~\ref{sbeyondHFB} and \ref{dbeyondHFB} illustrate, the energetic of doubly open-shell Cr isotopes is instead strongly impacted by the breaking of rotational symmetry. Indeed, most Cr isotopes do acquire a large intrinsic deformation\footnote{Interestingly, neutron deficient isotopes $^{34-42}$ are predicted to display a strong oblate-prolate oscillation. Isotopes between $N=20$ and $N=28$ all display a large prolate deformation, which slowly fades away towards $N=40$. Eventually, the prolate deformation suddenly increases again going across $N=40$ and stays large until the predicted neutron drip line at $N=48$.} as seen in the lower-right panel of Fig.~\ref{dbeyondHFB}. While the overall rms error of total binding energies remains similar in sBMBPT(2) and dBMBPT(2), the {\it evolution} with $N$ is strongly impacted as can be inferred from the behavior of $S_{2n}$ and $\Delta_{2n}$. 

As a matter of fact, the qualitative (quantitative) reproduction of $S_{2n}$ ($\Delta_{2n}$) is already excellent at the deformed mean-field level, i.e. all deficiencies identified in sHFB results are already corrected by dHFB. In particular, the fictitious shell closures at $N=20,28$ and $40$ have disappeared in dHFB results. Eventually, dynamical correlations added on top of dHFB via dBMBPT(2) increase $S_{2n}$ systematically to reach an excellent agreement with both VS-IMSRG(2) results and experimental data. While the rms error to experimental $S_{2n}$ was $2.9$\,MeV for sBMBPT(2) ($5.8$\,MeV for sHFB), it is $0.9$\,MeV for dBMBPT(2) ($4.4$\,MeV for dHFB), which is to be compared to $2.2$\,MeV for VS-IMSRG(2). 

Focusing on the 0f$_{7/2}$ shell, the right panels of Fig.~\ref{dbeyondHFBcurvature} show that the curvature of the energy is already very well captured at the dHFB level, while it was qualitatively wrong for both sHFB and sBMBPT(2), and becomes essentially as good as with VS-IMSRG(2) for dBMBPT(2). 

These results demonstrate that static correlations in doubly open-shell nuclei can be qualitatively and quantitatively seized via polynomially-scaling expansion methods built on top of a deformed  reference state and implemented to rather low truncation order.

\section{Sn chain}
\label{dMeanfieldsyst3}

\begin{figure*}
    \centering
    \includegraphics[width=0.7\textwidth]{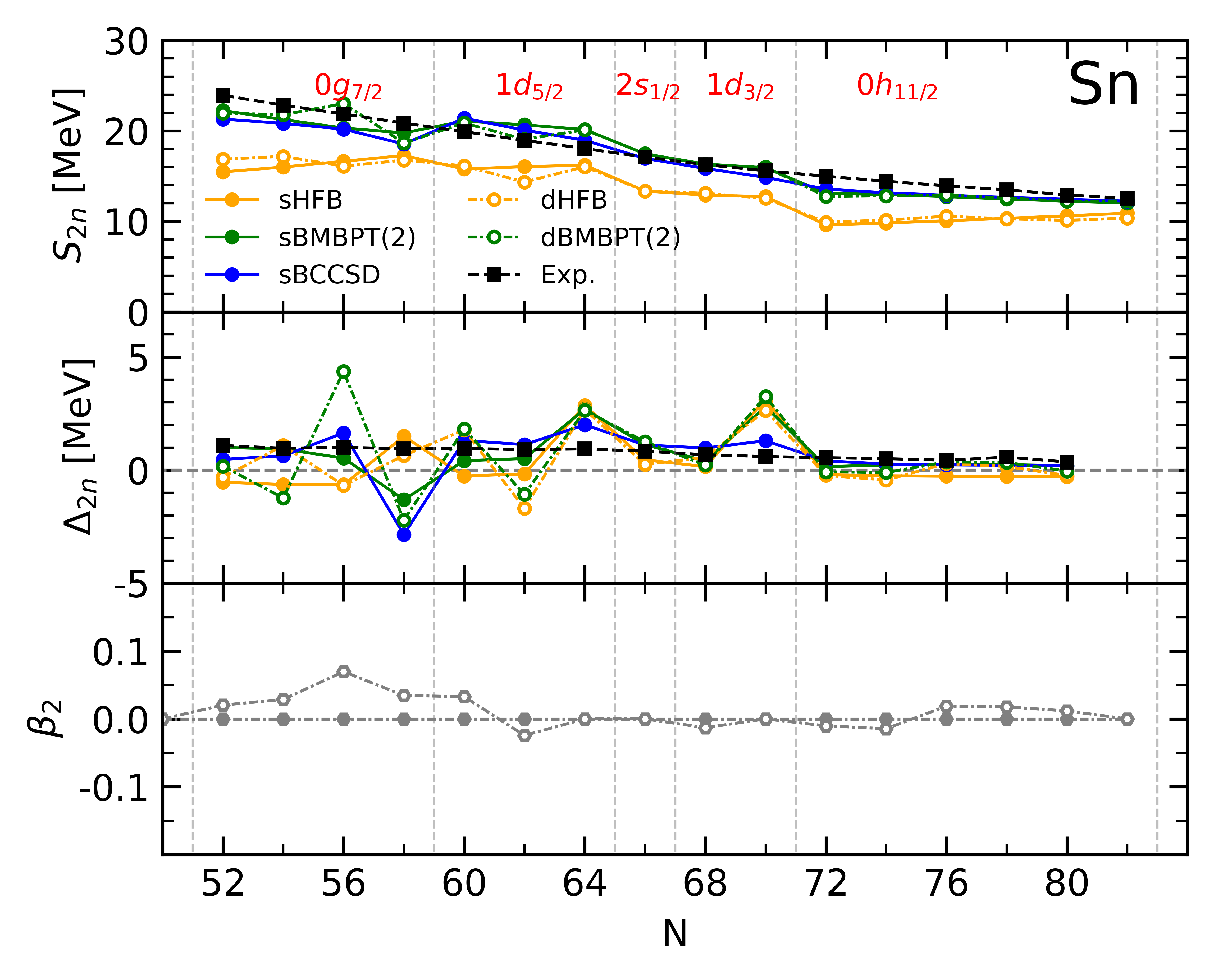}
    \caption{Results from sHFB, sBMBPT(2), sBCCSD, dHFB and dBMBPT(2) calculations against experimental data along the Sn isotopic chain. First line: two-neutron separation energy. Second line: two-neutron shell gap. Third line: intrinsic axial quadrupole deformation of sHFB and dHFB solutions.}
    \label{dbeyondHFBSn}
\end{figure*}

As a last step, the discussion is extended to semi-magic Sn isotopes between $^{100}$Sn and $^{132}$Sn, i.e. going through the sub-shell closures at $N=58, 64, 66$ and $70$ located between  the $N=50$ and $82$ major shell closures. In Fig.~\ref{dbeyondHFBSn}, $S_{2n}$ and $\Delta_{2n}$ computed from mean-field and beyond-mean-field calculations with and without breaking rotational symmetry are displayed. 

It is clear that experimental data do not show any fingerprint of the sub-shell closures, i.e. $S_{2n}$ decreases linearly between $N=52$ and $82$ such that $\Delta_{2n}$ is flat. Contrarily, sHFB results strongly reflect the presence of those sub-shell closures in a way that is consistent with the behavior seen in Ca isotopes, i.e. $S_{2n}$ are too low overall and rise linearly throughout open-shells, especially along the highly degenerate 0g$_{7/2}$ and 0h$_{11/2}$ shells.

Dynamical correlations brought on top of sHFB via sBMBPT(2) and sBCCSD largely ameliorate the situation, i.e. $S_{2n}$ are increased overall and the behavior throughout open-shells are corrected. However, the imprint of the sub-shell closures remain visible.

The larger mass combined with the weak pairing correlations induced by $\chi$EFT interactions at the mean-field level makes several semi-magic Sn isotopes take advantage of deformation if authorized to do so\footnote{This is again at variance with mean-field calculations based on effective EDFs. Indeed, the strong built-in pairing typically constrains all Sn isotopes to remain spherical between $N=50$ and $N=82$ in that case.} as can be seen from the lower panel of Fig.~\ref{dbeyondHFBSn}. Still, the axial quadrupole deformation parameter remains small in all cases. As in Ca isotopes, the wrong trend of $S_{2n}$ with $N$ observed at the sHFB level is thus not corrected by dHFB calculations and dBMBPT(2) eventually deliver very similar results to sBMBPT(2).

\section{Conclusions}
\label{conclusions} 

In order to extend the reach of \ai{} calculations to heavy doubly open-shell nuclei in the future, the most efficient strategy to incorporate dominant many-body correlations at play in (heavy) nuclei must be identified. With this in mind, the present work analyzed in details the impact of many-body correlations on binding energies of Calcium and Chromium isotopes with an (even) neutron number ranging from $N=12$ to $N=50$. 

Using an empirically-optimal (soft) $\chi$EFT-based Hamiltonian, binding energies computed in the spherical mean-field approximation were first shown to display specific shortcomings in semi-magic Ca isotopes. In addition to displaying a significant (but expected) underbinding, the corresponding energy was shown to evolve qualitatively incorrectly throughout (highly degenerate) open shells, i.e. whereas the linear decrease with the number of valence nucleons is too slow, the quadratic term makes the energy concave instead of being convex. Relying on the observation that $\chi$EFT-based interactions generate very little pairing at the spherical mean-field level, these two features could be related analytically to the fact that (i) associated single-particle energies are little bound compared to empirical one-nucleon separation energies and that (ii) the monopole valence-shell two-body matrix elements is attractive.  

Next, the consistent addition of dynamical correlations at polynomial cost via, e.g., low-order perturbation theory was shown to correct the deficiencies identified at the spherical mean-field level. This decisive improvement could also be understood analytically. Eventually, it is possible to reach a description of semi-magic Ca isotopes on essentially the same quantitative level as valence-space in-medium similarity renormalization group calculations, which rely on the diagonalization of the effective Hamiltonian in the $fp$ valence space. Either way, some yet missing correlation energy was identified between $^{40}$Ca and $^{48}$Ca that could be correlated with the (infamous) difficulty to describe the evolution of the charge radius between those two isotopes.

Moving to doubly open-shell Cr isotopes, calculations based on a spherical mean-field unperturbed state could not appropriately reproduce the binding energy evolution. However, allowing this unperturbed mean-field state to break rotational symmetry proved to be sufficient to capture the static correlations responsible for the phenomenological modifications observed between the two isotopic chains and that otherwise require the diagonalization of the effective Hamiltonian in large valence spaces.

Semi-magic Sn isotopes behave similarly to lighter Ca isotopes with a spherical mean-field delivering qualitatively wrong patterns that are corrected by the consistent addition of low-order dynamical correlations.

Eventually, the present work demonstrates that polynomially-scaling expansion methods based on unperturbed states possibly breaking (and restoring) symmetries constitute an optimal route to extend \ai{} calculations to heavy closed- and open-shell nuclei. To deepen the analysis and further consolidate the above conclusion, the present work will be extended to charge radii and $E2/E3$ transitions from the first $2^+/3^-$ excited states in the near future.

\section*{Acknowledgements}
The authors thank H. Hergert and T. Miyagi for providing the interaction matrix elements used in the numerical simulations. The work of A.S. was supported by the European Union’s Horizon 2020 research and innovation program under grant agreement No 800945 - NUMERICS - H2020-MSCA-COFUND-2017.
The work of P.D. was supported by the Research Foundation Flanders (FWO, Belgium, grant 11G5123N). The work of A.T. was supported by the European Research Council (ERC) under the European Union's Horizon 2020 research and innovation programme (Grant Agreement No.~101020842). The work of M.F. was carried out in the framework of the SINET project funded by the CEA.
Calculations were performed by using HPC resources from GENCI-TGCC (Contract No. A0150513012).

\section*{Data Availability Statement}
This manuscript has no associated data or the data will not be deposited.

\appendix

\section{Zero-pairing description}
\label{analytical}

The present work shows that $\chi$EFT-based interactions typically lead to a mean-field approximation displaying very weak pairing correlations in open-shell nuclei. This property is intimately linked to the fact that the total HFB  energy is concave rather than convex throughout long (enough) degenerate spherical shells. This connection can be validated analytically by considering that the system is in the extreme zero-pairing limit. 

The zero-pairing mean-field description of an open-shell system can be meaningfully achieved on the basis of two different many-body formalisms, i.e. the Hartree-Fock-Bogoliubov theory in the zero-pairing limit (HFB-ZP)~\cite{Duguet:2020hdm} or the Hartree-Fock theory in the equal filling approximation (HF-EFA)~\cite{Perez-Martin:2008dlm}. The two cases are worked out analytically below to validate the results obtained through realistic sHFB calculations in the body of the text.

\subsection{Hartree-Fock Bogoliubov}

The fully-paired HFB vacuum associated with a time-reversal symmetric system is written in its canonical, i.e., BCS-like, form as~\cite{RiSc80}
\begin{equation}
| \Phi \rangle \equiv \prod_{k>0} \left[u_k + v_k a^{\dagger}_k a^{\dagger}_{\bar{k}}\right] | 0 \rangle \, . \label{HFBstate}
\end{equation}
Operators $\{ a^{\dagger}_k, a_k\}$ characterize the so-called canonical one-body basis in which pairs of conjugate states $(k,\bar{k})$ are singled out by the Bogoliubov transformation via a quantum number $m_k$ such that $k\equiv (\breve{k},m_k)$ and $\bar{k}\equiv (\breve{k},-m_k)$. The state $\bar{k}$ ($k$) corresponds to the time-reversal state of $k$ ($\bar{k}$) up to a sign $\eta_{k}$ ($\eta_{\bar{k}}$) such that $\eta_{\bar{k}}\eta_{k}=-1$.

The BCS-like occupation numbers $u_k \equiv u_{\breve{k}}$ and $v_k= \eta_ {k} v_{\breve{k}}$ fulfilling $u_k^2+v_k^2=1$ are expressed in terms of the positive $m_k$-independent coefficients $(u_{\breve{k}},v_{\breve{k}})$. Employing the latter, the non-zero elements of the normal and anomalous density matrices read in the canonical basis as
\begin{align}
\rho_{kk}  =& \rho_{\bar{k}\bar{k}} = v^{2}_{\breve{k}} \, ,  \\
\kappa_{k\bar{k}}  =& -\kappa_{\bar{k}k} =  \eta_{k} u_{\breve{k}} v_{\breve{k}}  \, .
\end{align}

Based on the above and given the nuclear Hamiltonian 
\begin{align}
H  \equiv& \sum_{ij} t_{ij} a^{\dagger}_i  a_{j} \nonumber \\
&+ \frac{1}{(2!)^2} \sum_{ijkl} \overline{v}_{ijkl}  a^{\dagger}_i  a^{\dagger}_j a_l a_k \nonumber \\
&+ \frac{1}{(3!)^2}   \sum_{ijklmn} \overline{w}_{ijklmn}  a^{\dagger}_i  a^{\dagger}_j a^{\dagger}_k a_n a_m a_l \, ,
\end{align}
the total HFB energy reads in the canonical basis as
\begin{align}
E_\text{HFB}  \equiv& \langle \Phi | H | \Phi \rangle \nonumber \\
\equiv& E^{\text{kin}}_{| \Phi \rangle} + E^{\text{HF}}_{| \Phi \rangle} + E^{\text{B}}_{| \Phi \rangle} \nonumber \\
=& \sum_{k} t_{kk} \, v^2_{\breve{k}}   \nonumber \\
& + \frac{1}{2} \sum_{kk'} \overline{v}_{kk'kk'} \, v_{\breve{k}}^2  \, v_{\breve{k}'}^2  \nonumber \\
& + \frac{1}{4} \sum_{kk'} \overline{v}_{k\bar{k}k'\bar{k}'} \, \eta_{k} \eta_{k'} \, u_{\breve{k}}v_{\breve{k}} \, u_{\breve{k}'}v_{\breve{k}'}  \nonumber \\
&+ \frac{1}{6}  \sum_{kk'k''} \overline{w}_{kk'k''kk'k''}  \, v_{\breve{k}}^2  \, v_{\breve{k}'}^2  \, v_{\breve{k}''}^2   \nonumber \\
&+ \frac{1}{4} \sum_{kk'}\sum_{k''} \overline{w}_{k\bar{k}k''k'\bar{k}'k''} \, \eta_{k} \eta_{k'} \, u_{\breve{k}}v_{\breve{k}} \, u_{\breve{k}'}v_{\breve{k}'} \,  v_{\breve{k}''}^2   
\, . \label{HFBenergy}
\end{align}

Canonical single-particle states further gather in degenerate shells. All states belonging to a given shell share the same set of quantum numbers $\breve{k}$ and only differ by the value of $m_k$ such that the single-particle energy defining the shell is independent of it, i.e. $\epsilon_{k}=\epsilon_{\breve{k}}$. 

In the zero-pairing limit~\cite{Duguet:2020hdm}, states belonging to three categories of shells need to be distinguished according to
\begin{enumerate}
\item $\epsilon_{\breve{h}} - \lambda <0$, casually denoted as ``hole states",
\item $\epsilon_{\breve{v}} - \lambda =0$, casually denoted as ``valence states",
\item $\epsilon_{\breve{p}} - \lambda >0$, casually denoted as ``particle states" ,
\end{enumerate}
where $\lambda$ denotes the chemical potential. Accordingly, it can be shown that canonical states display the following average occupations
\begin{enumerate}
\item Hole state: $v^2_{\breve{h}} = 1$,
\item Valence state: $0< v^2_{\breve{v}} \leq 1$,
\item Particle state:  $v^2_{\breve{p}} = 0$.
\end{enumerate}
The valence shell gathers $p_{v}=d_{v}/2$ pairs of conjugated states such that the number of valence states $d_{v}$ (pairs $p_{v}$) is equal to the number of $m_v$ ($|m_v|$) different values. Consequently, the A nucleons making up the system are exhausted in such a way that $0\leq a_{v} \leq d_{v}$  of them sit in the valence shell whereas $A-a_{v}$ occupy the hole states. Consequently, the occupation of each of the $d_{v}$ valence states is
\begin{equation}
v^2_{\breve{v}} \equiv o_{\breve{v}} = \frac{a_{v}}{d_{v}} \, ,
\end{equation}
thus leading to
\begin{equation}
u_{\breve{v}}v_{\breve{v}} = \sqrt{o_{\breve{v}}(1-o_{\breve{v}})}  \, .
\end{equation}

Based on the above, the HFB energy (Eq.~\ref{HFBenergy}) of an open-shell system with $a_{v}$ nucleons in the valence shell can be computed relatively to the CS core in the zero-pairing limit.
After a lengthy but straightforward derivation, one obtains
\begin{align}
\Delta E^\text{HFB-ZP} (a_{v}) &\equiv  E^\text{HFB-ZP} (a_{v})- E^\text{HFB-ZP} (0)\\
&= \alpha_{\breve{v}} a_{v} + \frac{\beta_{\breve{v}}}{2} a_{v}^2 +   \frac{\gamma_{\breve{v}}}{6} a_{v}^3 \label{HFBZP}  \, , 
\end{align}
where 
\begin{subequations}
\label{derivatives} 
\begin{align}
\alpha_{\breve{v}} \equiv&   \epsilon^{\text{CS}}_{\breve{v}} + \frac{\Delta_{\breve{v}}}{4}   \, , \label{derivatives1} \\
\beta_{\breve{v}}  \equiv &  \frac{U_{\breve{v}}}{d_{v}}  -\frac{1}{2d_{v}}\left(\Delta_{\breve{v}} -Z_{\breve{v}}\right) \, , \label{derivatives2}  \\
\gamma_{\breve{v}}  \equiv &  \frac{1}{d^2_{v}} \left(X_{\breve{v}}-\frac{3}{2}Z_{\breve{v}}\right) \, , \label{derivatives3} 
\end{align}
\end{subequations}
where the valence-shell single-particle energy computed in the CS core
\begin{align}
\epsilon^{\text{CS}}_{\breve{v}} &\equiv t_{vv} + \sum_{h} \overline{v}_{vhvh} + \frac{1}{2} \sum_{hh'} \overline{w}_{vhh'vhh'} \, , \label{speCS}
\end{align}
and the $m_v$-independent quantities\footnote{The quantities introduced in Eq.~\ref{EffME} being independent of $m_v$, an additional sum over $m_v$ simply delivers a factor $d_{v}$.}
\begin{subequations}
\label{EffME} 
\begin{align}
U_{\breve{v}} & \equiv \sum_{m_{v'}}^{d_{v}} \left(\overline{v}_{vv'vv'} + \sum_{h} \overline{w}_{vv'hvv'h}\right) \nonumber \\
&\equiv \sum_{m_{v'}}^{d_{v}} \overline{\bold{v}}_{vv'vv'}  \, , \label{EffME2} \\
\Delta_{\breve{v}} &  \equiv  \eta_{v} \sum_{m_{v'}}^{d_{v}} \left(\overline{v}_{v\bar{v}v'\bar{v}'} + \sum_{h} \overline{w}_{v\bar{v}hv'\bar{v}'h}\right)  \eta_{v'} \nonumber \\
&\equiv \eta_{v} \sum_{m_{v'}}^{d_{v}} \overline{\bold{v}}_{v\bar{v}v'\bar{v}'}  \eta_{v'}\, , \label{EffME6} \\
X_{\breve{v}} &  \equiv  \sum_{m_{v'}}^{d_{v}} \sum_{m_{v''}}^{d_{v}} \overline{w}_{vv'v''vv'v''} \, , \label{EffME3} \\
Y_{\breve{v}} &  \equiv  \sum_{m_{v'}}^{d_{v}} \sum_{m_{v''}}^{d_{v}} \overline{w}_{vv'\bar{v}'vv''\bar{v}''} \, \eta_{v'} \eta_{v''} \, , \label{EffME4} \\
Z_{\breve{v}} &  \equiv   \eta_{v} \sum_{m_{v'}}^{d_{v}}  \sum_{m_{v''}}^{d_{v}} \overline{w}_{v\bar{v} v'' v'\bar{v}' v''} \,  \eta_{v'} \, ,\label{EffME7} 
\end{align}
\end{subequations}
have been introduced to express the results in a compact way. Equations~\eqref{EffME2} and~\eqref{EffME6} make use of the effective valence-shell two-body matrix elements 
\begin{align}
\overline{\bold{v}}_{vv'v''v'''}  &\equiv \overline{v}_{vv'v''v'''} + \sum_{h} \overline{w}_{vv'hv''v'''h}\, ,\label{EffME8} 
\end{align}
incorporating the contribution from the initial three-body interaction associated with an averaging over the CS core.

As demonstrated by Eqs.~\ref{HFBZP}-\ref{derivatives}, the HFB-ZP energy is manifestly\footnote{Equation~\ref{HFBZP} displays the explicit dependence of $\Delta E^\text{HFB-ZP}$ on $a_{v}$. However, additional implicit dependences are in fact at play in Eq.~\ref{HFBZP}. First, the two-body part of the center-of-mass kinetic energy correction included in the two-body interaction matrix elements actually depends on A. Second, all matrix elements at play carry an implicit dependence on $a_{v}$ through the nature of their indices. Indeed, canonical single-particle states are nucleus-dependent and thus evolve as the valence shell is being filled, i.e. with $a_{v}$. However, it was checked numerically that both effects are largely subleading.} cubic with the number of valence nucleons. While the cubic term originates entirely from the three-nucleon interaction, the curvature $\beta_{\breve{v}}$ of the HFB-ZP energy relates to a specific linear combination of two- and three-body matrix elements that can be extracted from actual HFB calculations. The sign of this combination of matrix elements determines the convexity or concavity character through the open shell, under the assumption that the cubic term is subleading, which can also be directly checked from a subset of three-body matrix elements.

The two-neutron separation energy of open-shell nuclei is given, for $a_v \geq 2$, by
\begin{align}
S^\text{HFB-ZP}_{2n}(a_{v}) \equiv& \Delta E^\text{HFB-ZP} (a_{v}\!-\!2) \!- \!\Delta E^\text{HFB-ZP} (a_{v}) \label{S2nHFBZP} \nonumber \\
=& \!-\!2(\alpha_{\breve{v}} \!-\!\beta_{\breve{v}} \!+\!\frac{2}{3} \gamma_{\breve{v}}) \!-\!2(\beta_{\breve{v}} \!- \!\gamma_{\breve{v}}) a_{v} \!-\! \gamma_{\breve{v}} \, a_{v}^2 \, . 
\end{align}
Under the (realistic) assumption that $|\alpha_{\breve{v}}| \gg |\beta_{\breve{v}}| \gg |\gamma_{\breve{v}}|$, $S_{2n}$ starts at $-2\alpha_{\breve{v}}$ and evolves linearly throughout the open shell with a negative (positive) slope $-2\beta_{\breve{v}}$ when the energy is convex (concave).

Following Eq.~\eqref{S2nHFBZP}, the two-neutron shell gap is given, for $a_v \geq 2$, by
\begin{align}
\Delta^\text{HFB-ZP}_{2n}(a_{v}) \equiv& S^\text{HFB-ZP}_{2n}(a_{v}) \!- S^\text{HFB-ZP}_{2n}(a_{v}) (a_{v}+2)  \nonumber \\
=& 4\beta_{\breve{v}} +4 \gamma_{\breve{v}} a_{v} \, . \label{Delta2nHFBZP}
\end{align}
Under the (realistic) assumption that $|\beta_{\breve{v}}| \gg |\gamma_{\breve{v}}|$, $\Delta_{2n}$ is constant throughout the open shell with a positive (negative) value when the energy is convex (concave).

Eventually, the evolution of the valence-shell single-particle energy as a function of $a_{v}$ is given by
\begin{align}
\epsilon^\text{HFB-ZP}_{\breve{v}}(a_{v}) =& \epsilon^{\text{CS}}_{\breve{v}} \nonumber \\
&+ 
\frac{1}{d_{v}}\left(U_{\breve{v}}  + \frac{1}{4} Y_{\breve{v}}  \right)a_{v} \nonumber \\ 
&+ \frac{1}{2d^2_{v}}\left(X_{\breve{v}}  -\frac{1}{2} Y_{\breve{v}} \right)a^2_{v} \, . \label{speHFBZP}
\end{align}
The valence-shell single-particle energy contains linear and quadratic contributions in $a_{v}$, the coefficient of the former (latter) being closely related to the curvature (cubic coefficient) of the HFB-ZP energy.

\subsection{Equal-filling approximation}

While the previous section provides analytical expressions derived within the frame of the HFB formalism in the zero-pairing limit~\cite{Duguet:2020hdm}, a simpler mean-field treatment of open-shell systems in absence of pairing correlations is provided by the HF theory in the equal filling approximation. While their results are closely related, the two formalisms are fundamentally different. Indeed, while HFB describes the system via a pure quantum state, the EFA is formulated within the frame of statistical quantum mechanics, i.e. the system is described in terms of a statistical density operator~\cite{Perez-Martin:2008dlm}. 

Effectively, EFA results can be trivially obtained by setting $\Delta_{\breve{v}}  =Y_{\breve{v}}  =Z_{\breve{v}}=0$ in the HFB-ZP formulae. Thus,  Eqs.~\eqref{HFBZP}, \eqref{S2nHFBZP} and \eqref{speHFBZP} apply, but with the modified coefficients
\begin{subequations}
\label{EFAderivatives} 
\begin{align}
\alpha_{\breve{v}} & =  \epsilon^{\text{CS}}_{\breve{v}}  \, , \label{EFAderivatives1} \\
\beta_{\breve{v}} & =  \frac{1}{d_{v}} U_{\breve{v}} \, , \\
\gamma_{\breve{v}} & =  \frac{1}{d^2_{v}} X_{\breve{v}}\, . \label{EFAderivatives3} 
\end{align}
\end{subequations}

\subsection{Discussion}

As already mentioned, numerical applications deliver $\gamma_{\breve{v}} = 0$ in all cases under scrutiny. Furthermore, the pairing contributions to $\alpha_{\breve{v}}$ and $\beta_{\breve{v}}$ are also negligible such that the HF-EFA results for $\gamma_{\breve{v}} = 0$ give an excellent account of HFB-ZP under the form
\begin{subequations}
\begin{align} 
\Delta E^\text{HF-EFA}(a_{v}) &= \epsilon^{\text{CS}}_{\breve{v}}  a_v + \frac{\beta_{\tilde{v}}}{2} a^2_{v} \, ,  \\
S^\text{HF-EFA}_{2n}(a_{v})&= -2 \epsilon^{\text{CS}}_{\breve{v}}   - 2\beta_{\tilde{v}} (a_{v} -1)  \, , \\
\epsilon^\text{HF-EFA}_{\tilde{v}}(a_{v}) &= \epsilon^{\text{CS}}_{\breve{v}} + \beta_{\tilde{v}} a_{v}\, , \\
\Delta^\text{HF-EFA}_{\tilde{v}}(a_{v}) &= 4\beta_{\tilde{v}} \, .
\end{align}
\end{subequations}
The evolutions of the total binding energy, the two-nucleon separation and the valence-shell single-particle energy, as one fills the valence shell, are strictly correlated and entirely driven by the valence-shell single-particle energy computed in the core $\epsilon^{\text{CS}}_{\breve{v}}$ (diagrammatically represented in Fig.~\ref{diagrams1storder_e}) and by $\beta_{\breve{v}}$ that is nothing but the average diagonal matrix elements of the effective valence-shell two-body interaction (diagrammatically represented in Fig.~\ref{diagrams1storder_v})
\begin{align} 
\beta_{\breve{v}} &= \frac{1}{d_{v}} \sum_{m_{v'}}^{d_{v}} \overline{\bold{v}}_{vv'vv'}  \, .
\end{align}

More specifically, while the total energy is quadratic in $a_{v}$, the two-nucleon separation energy and the valence-shell single-particle energy are linear. The coefficient of the linear (quadratic) term in the former drives the initial value (slope) of the latter, knowing that the slopes of the two-nucleon separation energy and of single-particle energy are opposite.

\subsection{Second-order MBPT}
\label{MBPT2app}

Having semi-analytical expressions as a function of $a_v$ for the mean-field results in the zero-pairing limit, it is now relevant to investigate the addition of dynamical correlations. 

This is presently done by evaluating the MBPT(2) corrections to the valence-shell single-particle energy computed in the CS core and to the valence-shell effective two-body interaction. To do so, the valence shell is taken to be doubly degenerate $(v,v')$\footnote{This setting is mandatory to compute the energy of two successive even isotopes via MBPT, i.e. to avoid actually dealing with open-shell systems given that pairing was shown to be negligible for the present discussion and given that no perturbation theory based on a HF-EFA statistical operator is available to date.} and the total energy is computed at the MBPT(2) level for both the CS core and the system with two more particles in order to compute the two-nucleon separation energy.  

After a lengthy but straightforward derivation, the separation energy between the two even isotopes is obtained as
\begin{align} 
S^\text{(2)}_{2n}(2) &= E^{(2)}(0) - E^{(2)}(2) \nonumber \\
&= -2\left(\epsilon^{\text{CS}}_{\breve{v}} + \Sigma^{(2)}_{\breve{v}}(\epsilon^{\text{CS}}_{\breve{v}}) \right)  \nonumber \\
&\hspace{0.4cm} - \left(\overline{\bold{v}}_{vv'vv'} + \overline{\bold{v}}^{(2)}_{vv'vv'}(\epsilon^{\text{CS}}_{\breve{v}})\right)  \, ,
\end{align}
where the (on-shell) valence-shell self-energy and two-body effective interaction corrections are given by
\begin{subequations}
\label{corrections2ndorder}
\begin{align} 
\Sigma^{(2)}_{\breve{v}}(\epsilon^{\text{CS}}_{\breve{v}}) &= +\frac{1}{2} \sum_{hh'p} \frac{|\overline{\bold{v}}_{hh'vp}|^2}{\epsilon^{\text{CS}}_{p}+\epsilon^{\text{CS}}_{\breve{v}}-\epsilon^{\text{CS}}_{h}-\epsilon^{\text{CS}}_{h'}} \nonumber \\
&\hspace{0.4cm} - \frac{1}{2} \sum_{pp'h} \frac{|\overline{\bold{v}}_{vhpp'}|^2}{\epsilon^{\text{CS}}_{p}+\epsilon^{\text{CS}}_{p'}-\epsilon^{\text{CS}}_{h}-\epsilon^{\text{CS}}_{\breve{v}}}\, , \\
\overline{\bold{v}}^{(2)}_{vv'vv'}(\epsilon^{\text{CS}}_{\breve{v}}) &=  +\frac{1}{2} \sum_{hh'}  \frac{|\overline{\bold{v}}_{hh'vv'}|^2}{2\epsilon^{\text{CS}}_{\breve{v}}-\epsilon^{\text{CS}}_{h}-\epsilon^{\text{CS}}_{h'}} \nonumber \\
&\hspace{0.4cm} - \frac{1}{2} \sum_{pp'} \frac{|\overline{\bold{v}}_{vv'pp'}|^2}{\epsilon^{\text{CS}}_{p}+\epsilon^{\text{CS}}_{p'}-2\epsilon^{\text{CS}}_{\breve{v}}} \, ,
\end{align}
\end{subequations}
and displayed diagrammatically in Figs~\ref{diagrams2ndorder_e} and~\ref{diagrams2ndorder_v}, respectively. The second-order corrections to the total binding energy translate for the two-neutron separation energy into a correction of the mean-field valence-shell single-particle energy and of the effective valence-shell two-body interaction. 

Extending candidly the situation to a $d_{v}$-fold degenerate valence-shell in a EFA-like spirit, $S_{2n}$ and $\Delta_{2n}$ evolve for $a_v \geq 2$ as 
\begin{subequations}
\label{s2n2ndorder}
\begin{align} 
S^\text{(2)}_{2n}(a_{v})&= -2\alpha^{(2)}_{\tilde{v}} - 2\beta^{(2)}_{\tilde{v}} (a_{v} -1)  \, , \\
\Delta^\text{(2)}_{2n}(a_{v})&= 4\beta^{(2)}_{\tilde{v}}  \, , 
\end{align}
\end{subequations}
with 
\begin{subequations}
\label{effint2ndorder}
\begin{align} 
\alpha^{(2)}_{\tilde{v}} &\equiv \epsilon^{\text{CS}}_{\breve{v}} + \Sigma^{(2)}_{\breve{v}}(\epsilon^{\text{CS}}_{\breve{v}}) \, , \\
\beta^{(2)}_{\tilde{v}} &\equiv \frac{1}{d_{v}} \sum_{m_{v'}}^{d_{v}} \left(\overline{\bold{v}}_{vv'vv'} + \overline{\bold{v}}^{(2)}_{vv'vv'}(\epsilon^{\text{CS}}_{\breve{v}})\right) \, , 
\end{align}
\end{subequations}
the latter being the averaged valence-shell interaction at second order in perturbation theory.

As seen in Eq.~\eqref{s2n2ndorder} and~\eqref{effint2ndorder}, dynamical correlations modify both the starting value of the $S^\text{(2))}_{2n}$ in the valence-shell and the slope governing its evolution, i.e. the self-energy correction impacts the former whereas the correction to the valence-shell interaction modifies the latter.

\section{$\Delta_{2n}$}
\label{app2ndderivative}

The two-neutron shell gap defined in Eq.~\eqref{eq:Delta2n} explicitly reads as
\begin{equation}
\Delta_{2n}(N, Z) 
= E(N-2, Z) - 2E(N, Z) + E(N+2, Z).
\label{eqn:d2n_proof}
\end{equation}
The second derivative of the total energy centered around $N$ can be written through finite difference coefficients as
\begin{equation}
\dfrac{\partial^2 E(N, Z)}{\partial N^2} = \dfrac{1}{4} (E(N-2, Z) - 2E(N, Z) + E(N+2, Z)) \, ,
\end{equation}
which proves that
\begin{equation}
\dfrac{\partial^2 E(N, Z)}{\partial N^2} = \dfrac{\Delta_{2n}(N, Z)}{4} \, .
\end{equation}

\bibliography{bibliography.bib}

\end{document}